\documentclass[a4paper]{article}

\usepackage[english]{babel}
\usepackage[utf8x]{inputenc}
\usepackage[T1]{fontenc}
\usepackage{authblk}

\usepackage[a4paper,top=3cm,bottom=2cm,left=3cm,right=3cm,marginparwidth=1.75cm]{geometry}

\usepackage{multirow}
\usepackage{amsmath}
\usepackage{graphicx}
\usepackage{hyperref}
\hypersetup{colorlinks = true, allcolors = blue}
\title{
Circular and Linear ${\rm e^+ e^-}$ Colliders:\\ Another Story of Complementarity\\
{\small \it Contribution to the European Strategy for Particle Physics Update, 2018--2020}
}
\author[$1$]{Alain Blondel\thanks{Alain.Blondel@cern.ch}}
\author[$2$]{Patrick Janot\thanks{Patrick.Janot@cern.ch}}
\affil[1]{\small LPNHE, Sorbonne Universit\'e, 4 Place Jussieu, 75252 Paris, France}
\affil[2]{\small CERN, EP Department, 1 Esplanade des Particules, CH-1217 Meyrin, Switzerland}

\date{\small 8 January 2020}

\begin{document}
\maketitle

\begin{abstract}
The remarkable synergy and complementarity between the circular ${\rm e^+ e^-}$ and pp colliders has been extensively discussed. In this short document, we investigate the  complementarity between the proposed circular and linear ${\rm e^+ e^-}$ colliders at the electroweak and TeV scale.  This complementarity could be exploited on a world-wide scale, if both a large circular and a linear infrastructures were available. A possible implementation of such a complementary program is shown.  

\end{abstract}

\vfill\eject
\tableofcontents
\vfill\eject
\section{Introduction} 
 
One of the main topics being considered by the ongoing European Strategy for Particle Physics (ESPP) update is the ambition to recommend a clear direction for the preparation and the technical study of an ambitious post-LHC infrastructure at CERN, in view of a final decision at the following strategy process. The question is ``Which one?'' and more precisely ``Should this infrastructure be circular or linear?''. 

Two elements of consensus emerge from the  presentations and discussions that took place in particular at the public symposium in Granada and in the subsequent Physics Briefing Book~\cite{Heinemann:2691414}. 

\begin{enumerate}
    \item There is a strong physics case for an $\rm e^+ e^-$ collider to measure the Higgs boson and other particle properties. In particular, the ${\rm e^+ e^- \rightarrow Z H}$ and ${\rm e^+ e^- \rightarrow \rm t \bar t}$ processes allow the Higgs and top couplings to the Z boson and  the  total  Higgs boson width to be extracted in a model-independent and absolute way. With these essential inputs, all hadron collider Higgs measurements can in turn be rendered absolute by normalisation to the $\rm H \rightarrow  ZZ $ decay and the ${\rm pp \to t\bar t Z}$ cross section, used as  ``fixed candles''.
    \item The highest elementary parton-parton collision energy can be achieved, for the foreseeable future, with a high-energy proton-proton collider, for which a circular geometry is the only available option, at least for energies up to $\approx$ 150\,TeV in a ring of 100\,km circumference.   
    
\end{enumerate}

The FCC study ultimately aims at 100\,TeV pp collisions and possibly beyond. Having considered a number of alternatives, it concluded that the full FCC integrated programme (FCC-INT~\cite{Benedikt:2653673}), which features in sequence a high-luminosity $\rm e^+ e^-$ electroweak, flavour, Higgs, and top factory (FCC-ee), followed by a $ \geq 100$\,TeV pp collider (FCC-hh), is the most pragmatic, most effective, and safest implementation of a global HEP research infrastructure addressing the above two points. This scenario is similar to the LEP/LHC sequence. It offers a broad and powerful program, well adapted to the physics landscape, in which we know that new physics must exist, whose nature, couplings, and energy scale are unknown. In this ambitious perspective, CERN offers unique assets: existing powerful infrastructure; outstanding personnel expertise, built over several decades; long-term budget stability; all being  essential prerequisites for such a large-scale project. 

The FCC-INT scenario is also well matched to the large community of CERN users, and would maintain Europe in its leading role in fundamental physics and in the related technologies. To ensure that FCC can start operation seamlessly at the end of HL-LHC with the lepton collider FCC-ee as first step, a clear orientation towards the circular option is required in this round of the strategy process. The development of efficient and affordable key technology for FCC-hh, ready for mass production, will require several decades, fitting well the overall FCC planning. Alternative first steps, such as the HE-LHC or a low-energy 100\,km hadron collider (LE-FCC), end up being more expensive and less physics-rich than the FCC-INT scenario\footnote{The FCC International Advisory Committee reviewed the issue in its meeting of October 2019~\cite{IAC-2019-10} and concluded: {\em The available cost estimate [of the LE-FCC option] is still high, especially in view of the limited physics reach. In a staging scenario, it is not attractive to replace the FCC-ee option by the low-energy proton version.}}, and consequently risk to jeopardise the whole programme. 

A difficulty in the strategy process appears to be that more than one $\rm e^+ e^-$ facility satisfies the requirement (1) above, while it is generally assumed that only one can go forward. In addition, by fulfilling both requirements (1) and (2), the FCC-INT program might be perceived as a threat that automatically eliminates all other proposals, with the risk that CERN might remain alone on the global map of the high-energy frontier. 

The most advanced project at the high-energy frontier outside of the CERN domain is the International Linear Collider (ILC), presently proposed by the Linear Collider Collaboration (LCC) for consideration by the Japanese Government. Linear colliders have been considered to be the path to high-energy ${\rm e^+ e^-}$ collisions for more than 40 years~\cite{Amaldi:1975hi}. A considerable design effort has led to the presentation of a TDR~\cite{Adolphsen:2013jya,Adolphsen:2013kya} in 2013, as well as significant luminosity upgrades for the ESPP 2018-2020~\cite{Bambade:2019fyw}. Is ILC made redundant by FCC-ee, and vice versa? Certainly, the members of the LCC (resp. FCC) are keen on putting forward that, should FCC-ee (resp. ILC) go ahead without ILC (resp. FCC-ee), considerable physics opportunities might be lost.   

This note is organised as follows. The complementarity is primarily contained in the respective collider performance, briefly reviewed in  Section~\ref{sec:accelerator}. We then proceed in Section~\ref{sec:scientific} to a non-exhaustive discussion of the complementarity in various domains of the physics program: Higgs properties (\ref{sec:Higgs}), Electroweak measurements (\ref{sec:EW}), Flavour and QCD aspects (\ref{sec:Flavour}), top-quark physics (\ref{sec:Top}), and searches beyond the Standard Model (\ref{sec:BSM}). A summary of the physics complementarity is tabulated in Section~\ref{sec:summary}. Possible cost optimisations, which can be contemplated if both machines proceeded to realisation, are discussed in Section~\ref{sec:financial}. Sociological considerations are examined in Section~\ref{sec:sociological}.  A final comment on a global vision is offered in Section~\ref{sec:regional}. Because global resources are limited, it is difficult to benefit simultaneously from all the opportunities offered by operational, scientific, financial, sociological, and regional complementarity. A possibly realistic implementation of such a complementary program, which tentatively emphasises the regional, financial, and operational aspects, and maximises the scientific outcome, is discussed in Section~\ref{sec:scenario}.

\section{Complementarity of collider performance}
\label{sec:accelerator}

\begin{figure}[htbp]
\centering
\includegraphics[width=0.49\textwidth]{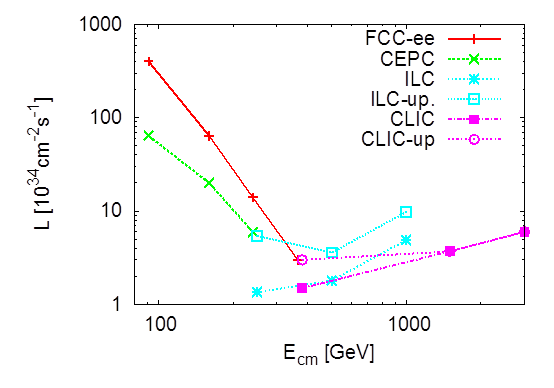}
\includegraphics[width=0.49\textwidth]{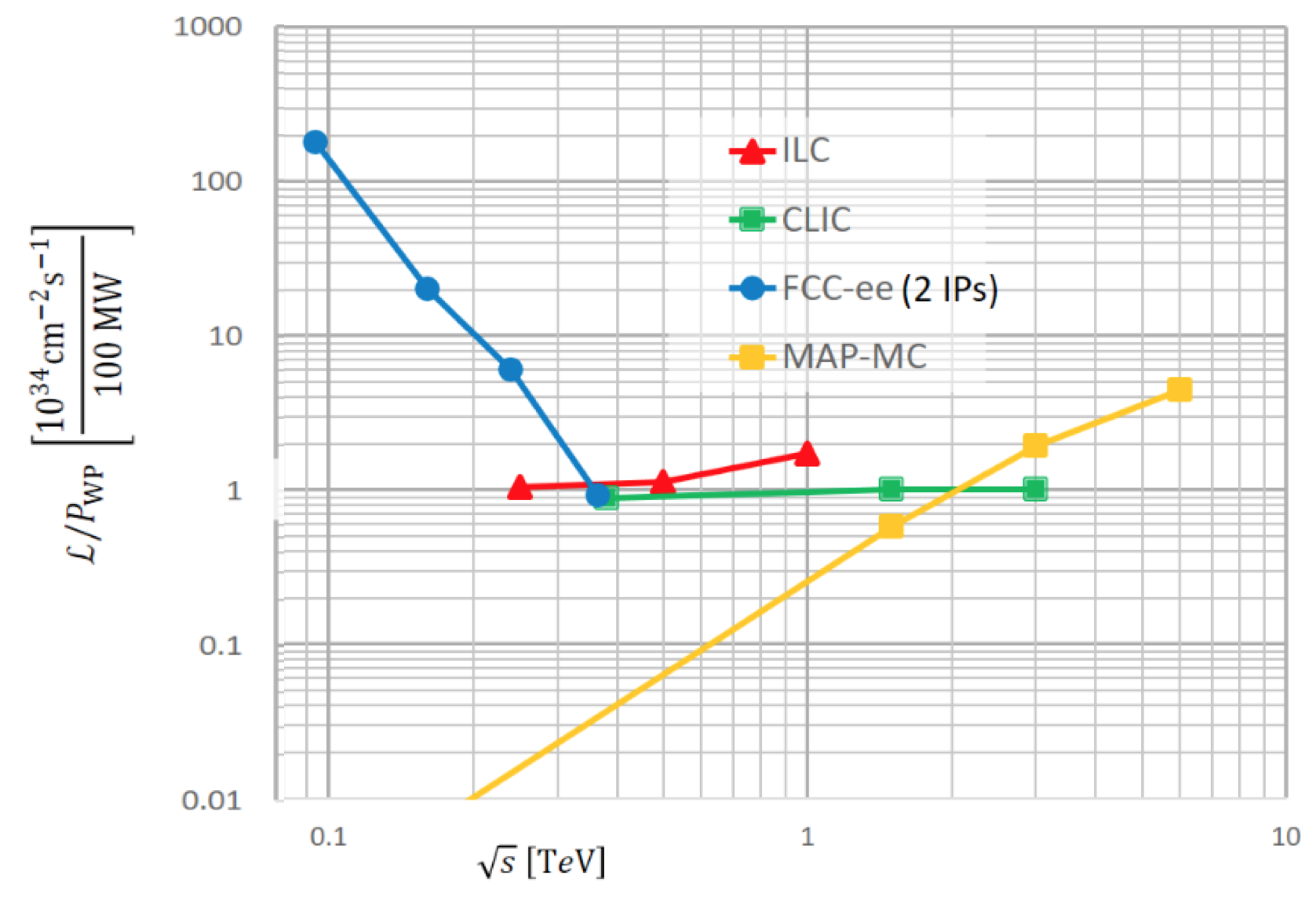}
\caption{\label{fig:luminosities} \small Luminosity (left) and luminosity per power (right), as a function of the centre-of-mass energy for the proposed ${\rm e^+e^-}$ colliders at the electroweak scale and their potential upgrades. Two interaction points (IPs) are assumed for the circular colliders FCC-ee and CEPC. From the Physics Briefing Book~\cite{Heinemann:2691414}.}
\end{figure}

The operational complementarity is best illustrated in Fig.~\ref{fig:luminosities}, which displays the luminosity, and the luminosity per total facility power, expected to be produced at the various ${\rm e^+e^-}$ collider projects, as a function of the centre-of-mass energy. 

The circular colliders at the electroweak scale (FCC-ee and CEPC) were conceived in 2011--2013, as soon as the Higgs boson became known to be light~\cite{Blondel:2011fua}. Their luminosity curves provide the highest statistics at low energies, but are strongly limited by synchrotron radiation above 350--400\,GeV. The proposed operation models comprise data taking at and around the Z pole, at the WW threshold, at the ZH cross section maximum (240 GeV), and for FCC-ee, an extension up to 365 GeV at and above the top pair threshold. The designs are sufficiently flexible to allow for operation at other centre-of-mass energies, if justified by compelling physics arguments (e.g., at $\sqrt{s} = m_{\rm H}$, or well below the Z peak), with unrivalled luminosities. Both colliders are planned to operate for 10--15 years (of which 7--8 years at 240\,GeV and above)  with two IPs\footnote{A configuration with 4 IPs is being studied for FCC-ee, allowing the total luminosity to increase by a factor 1.7 with little increase of total power.}, and are considered to be an enabling first step in a long-term plan towards a high-energy proton-proton collider delivering the highest parton-parton centre-of-mass energies. Such a hadron collider would then operate for two or three decades.

The linear colliders (ILC and CLIC) have been studied since 1975~\cite{Amaldi:1975hi}, and are considered to be the only possible way towards high-energy ${\rm e^+e^-}$ collisions. Luminosity and power consumption grow linearly with energy. The proposed operation models include a first run at ``low'' energy, 250\,GeV for ILC~\cite{Evans:2017rvt,Bambade:2019fyw} and 380\,GeV for CLIC~\cite{CLIC:2016zwp}, for about a decade, with longitudinal beam polarisation. Both colliders have an open-ended run plan, with possible upgrades to 1\,TeV (ILC) and 3\,TeV (CLIC) in operation models that extends over several decades. 

At the top-pair threshold (and immediately above), FCC-ee and linear colliders are planned to deliver similar integrated luminosities, within a factor of two. A linear collider is the most effective option at 500\,GeV (and the only possibility above), while a circular collider wins for any energy below 350\,GeV. For example, at the ZH cross-section maximum, FCC-ee is expected to produce $5\,{\rm ab}^{-1}$ in about three years, while it would take between 20 and 30 years with ILC to reach the same figure; the integrated power for the circular machine is also 5-10 times less per Higgs boson produced in this energy domain (right panel of Fig.~\ref{fig:luminosities}). The best return-on-investment plan would therefore be to explore the lower energy range (up to 365\,GeV) with a circular collider, and the higher energy range from 250\,GeV upwards with a linear collider. 

\section{Scientific complementarity}
\label{sec:scientific}
\subsection{Higgs physics}
\label{sec:Higgs}

\begin{figure}[htbp]
\centering
\includegraphics[width=0.44\textwidth]{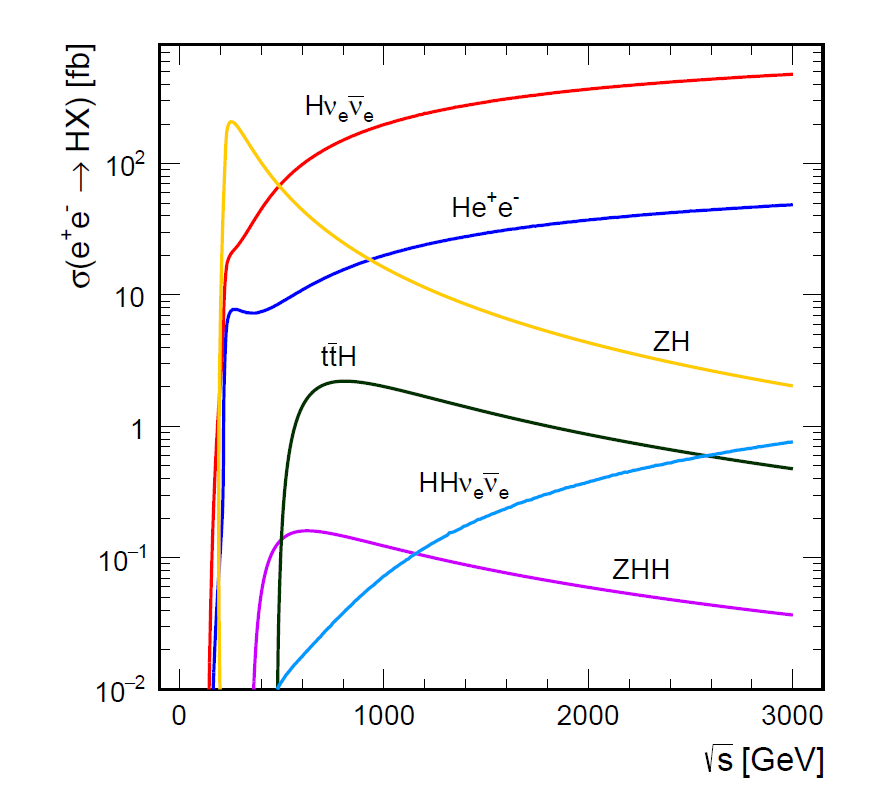}
\includegraphics[width=0.54\textwidth]{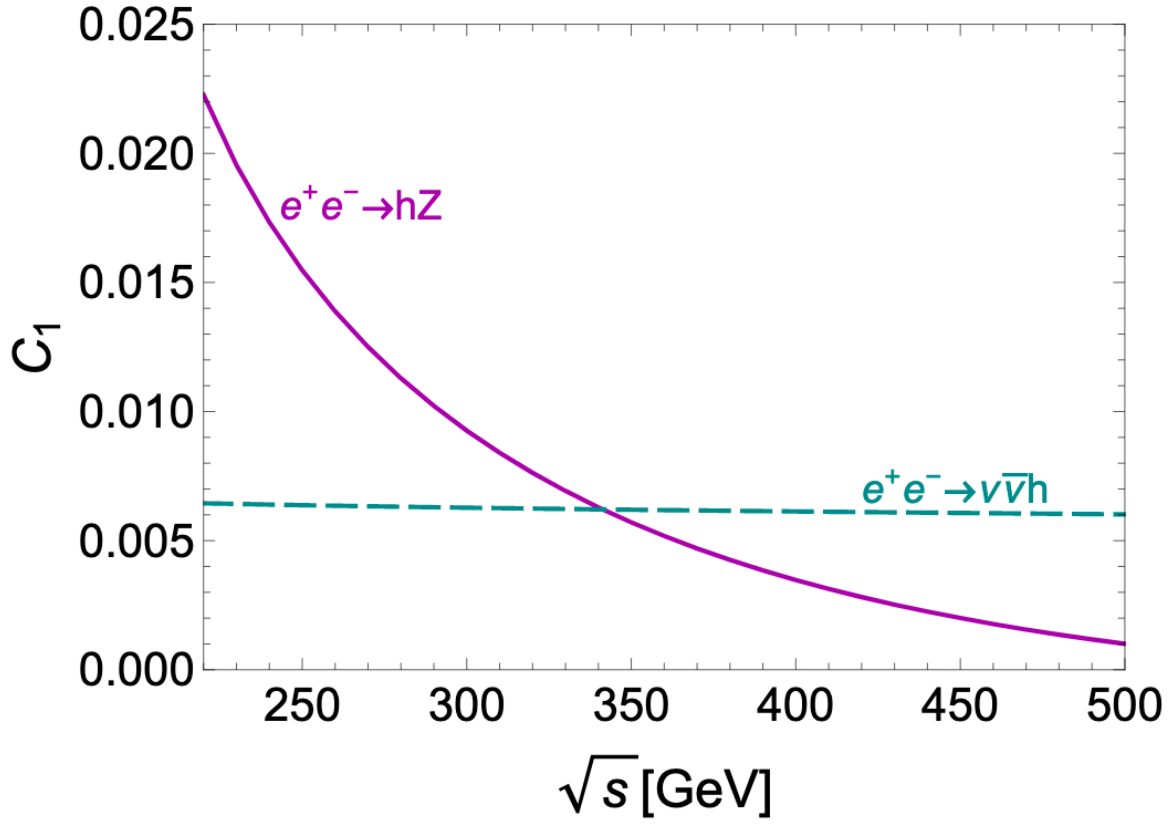}
\caption{\small (Left) Higgs production cross sections as function of centre-of-mass energy, from Ref.~\cite{Heinemann:2691414}. (Right) Relative enhancement of the ZH and ${\rm H\nu_e\bar\nu_e}$ cross sections due to the Higgs self-coupling appearing in one-loop SM diagrams, from Ref.~\cite{DiVita:2017vrr}.}
\label{fig:Hsec}
\end{figure}

After completion of their proposed operation models (up to 365\,GeV for FCC-ee, up to 1\,TeV for ILC, and up to 3\,TeV for CLIC), a substantial part of the $\rm e^+e^-$ collider Higgs physics programs is similar. There are, however, significant differences due to the variation of the production mode  as a function of energy, as can be seen in the left panel of Fig.~\ref{fig:Hsec}. 
A summary of the measurement capabilities of the various colliders is given in Fig.~3.8 of the Physics Briefing Book~\cite{Heinemann:2691414}, and compiled in Table~\ref{tab:kappaEFT} for the low-energy Higgs factories, and in Table~\ref{tab:EnergyUpgrades} for their combination with the corresponding high-energy colliders.

{\setlength{\tabcolsep}{6pt} 
\renewcommand{\arraystretch}{1.} 
\begin{table}[!htbp]
\centering
\caption{\small Precision on the Higgs boson couplings, from Ref.~\protect\cite{deBlas:2019rxi}, in the $\kappa$ framework without (first numbers) and with (right numbers) a combination with HL-LHC projections~\protect\cite{Cepeda:2019klc}, for the five low-energy Higgs factories (ILC$_{250}$, CLIC$_{380}$, CEPC$_{240}$, and FCC-ee$_{240\to 365}$ with 2 IPs). For $g_{\rm HHH}$, the result of a global EFT fit is shown with 2 IPs (top) and 4 IPs (bottom). All numbers are in \% and indicate 68\% C.L. sensitivities. Also indicated are the standalone precision on the total decay width and the 95\% C.L. sensitivity on the "invisible" and "exotic" branching fractions, the latter accounting for final states that cannot be tagged as SM decays. All numbers include current projected parametric uncertainties. The HL-LHC result is obtained by fixing the total Higgs boson width and the ${\rm H \to c\bar c}$ branching fraction to their Standard Model values, and by assuming no BSM decays. The CEPC team has shown that a significant result for the HZ$\gamma$ coupling can be achieved from the large sample of Higgs bosons accessible at circular ${\rm e^+e^-}$ colliders. The HZ$\gamma$ coupling is otherwise obtained solely from HL-LHC projections. A result similar to that obtained with CEPC can be expected from FCC-ee.}
\label{tab:kappaEFT}

\vspace{2mm}
\begin{tabular}{|l|c|c|c|c|c|}
\hline Collider & {HL-LHC} &  { ILC$_{250}$} & { CLIC$_{380}$} & { CEPC$_{240}$} & FCC-ee$_{240\to 365}$ \\ \hline
Lumi (${\rm ab}^{-1}$) &  3 & 2 & 1 & 5.6 &  5 + 0.2 + 1.5 \\ \hline
Years & { 10} & 11.5 & 8 & 7 & 3 + 1 + 4 \\ \hline
$g_{\rm HZZ}$ (\%) &  1.5  & 0.30 / 0.29 & 0.50 / 0.44  & 0.19 / 0.18 & 0.18 / 0.17  \\ 
$g_{\rm HWW}$ (\%) &  1.7  & 1.8 / 1.0 & 0.86 / 0.73 & 1.3 / 0.88 & 0.44 / 0.41  \\ 
$g_{\rm Hbb}$ (\%) &  5.1 & 1.8 / 1.1 & 1.9 / 1.2 & 1.3 / 0.92 & 0.69 / 0.64 \\ 
$g_{\rm Hcc}$ (\%) &  SM & 2.5 / 2.0 & 4.4 / 4.1 & 2.2 / 2.0 & 1.3 / 1.3 \\ 
$g_{\rm Hgg}$ (\%) &  2.5  & 2.3 / 1.4 & 2.5 / 1.5 & 1.5 / 1.0 & 1.0 / 0.89 \\ 
$g_{\rm H\tau\tau}$ (\%) &  1.9 & 1.9 / 1.1 & 3.1 / 1.4 & 1.4 / 0.91 & 0.74 / 0.66 \\ 
$g_{\rm H\mu\mu}$ (\%) &  4.4  & 15. / 4.2 & -- / 4.4 & 9.0 / 3.9 & 8.9 / 3.9 \\ 
$g_{\rm H\gamma\gamma}$ (\%) &  1.8  & 6.8 / 1.3 & -- / 1.5 & 3.7 / 1.2 & 3.9 / 1.2 \\ 
$g_{\rm HZ\gamma}$ (\%) &  11.  & -- / 10. & -- / 10.  & 8.2 / 6.3 & -- / 10. \\ 
$g_{\rm Htt}$ (\%) &  3.4 & -- / 3.1 & -- / 3.2 & -- / 3.1 & 10. / 3.1 \\ \hline
\multirow{2}{*}{$g_{\rm HHH}$ (\%)} &  \multirow{2}{*}{50.} &  \multirow{2}{*}{ -- / 49.} & \multirow{2}{*}{ -- / 50. } & \multirow{2}{*}{ -- / 50. } & 44./33. \\ 
& & & & & {27./24.} \\ \hline
$\Gamma_{\rm H}$ (\%) &  SM & 2.2 & 2.5 & 1.7 & 1.1  \\ \hline
BR$_{\rm inv}$ (\%) &  1.9 &  0.26 & 0.65 & 0.28 & 0.19 \\ 
BR$_{\rm EXO}$ (\%) &  SM (0.0) & 1.8 & 2.7 & 1.1 & 1.1 \\ \hline
\end{tabular} 
\end{table}
}

{\setlength{\tabcolsep}{6pt} 
\renewcommand{\arraystretch}{1.} 
\begin{table}[p]
\centering
\caption{\small Precision on the Higgs boson couplings from Ref.~\protect\cite{deBlas:2019rxi}, in the $\kappa$ framework without (first numbers) and with (second numbers) HL-LHC projections, for the combination of each low-energy Higgs factory (ILC$_{250}$, CLIC$_{380}$, and FCC-ee) and their proposed upgrades: ILC$_{500\,{\rm GeV}}$, ILC$_{1000\,{\rm GeV}}$,  CLIC$_{1.4+3\,{\rm TeV}}$, and the complete FCC integrated programme. All numbers are in \% and indicate 68\% C.L. sensitivities. Also indicated are the standalone precision on the total decay width, and the 95\% C.L. sensitivity on the "invisible" and "exotic" branching fractions. A precision similar to that achieved at high-energy linear colliders is reached with FCC-hh in less than one year of operation for all couplings, except the Higgs self-coupling for which a precision of 10\% is reached in about 7 years (with respect to a couple decades for ILC$_{1000}$ and CLIC.)} 
\label{tab:EnergyUpgrades} 

\vspace{2mm}
\begin{tabular}{|l|c|c|c|c|}
\hline Collider & {ILC$_{500}$} & {ILC$_{1000}$} & {CLIC} & FCC-INT \\ \hline
$g_{\rm HZZ}$ (\%) & 0.24 / 0.23   & 0.24 / 0.23 & 0.39 / 0.39 & 
0.17 / 0.16 \\ 
$g_{\rm HWW}$ (\%) & 0.31 / 0.29 & 0.26 / 0.24 & 0.38 / 0.38 & 0.20 / 0.19 \\ 
$g_{\rm Hbb}$ (\%) & 0.60 / 0.56 & 0.50 / 0.47  & 0.53 / 0.53  & 0.48 / 0.48 \\ 
$g_{\rm Hcc}$ (\%) & 1.3 / 1.2 & 0.91 / 0.90 & 1.4 / 1.4 & 0.96 / 0.96 \\ 
$g_{\rm Hgg}$ (\%) & 0.98 / 0.85 & 0.67 / 0.63 & 0.96 / 0.86  & 0.52 / 0.50\\ 
$g_{\rm H\tau\tau}$ (\%) & 0.72 / 0.64 & 0.58 / 0.54 & 0.95 / 0.82 & 0.49 / 0.46\\ 
$g_{\rm H\mu\mu}$ (\%) & 9.4 / 3.9 & 6.3 / 3.6  & 5.9 / 3.5 & 0.43 / 0.43 \\ 
$g_{\rm H\gamma\gamma}$ (\%) & 3.5 / 1.2 & 1.9 / 1.1 & 2.3 / 1.1 & 0.32 / 0.32 \\ 
$g_{\rm HZ\gamma}$ (\%) & -- / 10.  & -- / 10. & 7. / 5.7 & 0.71 / 0.70\\ 
$g_{\rm Htt}$ (\%) & 6.9 / 2.8 & 1.6 / 1.4 & 2.7 / 2.1 &  1.0 / 0.95 \\ \hline
$g_{\rm HHH}$ (\%) & 27. & 10. & 9. & 5. \\ \hline
$\Gamma_{\rm H}$ (\%) & 1.1 & 1.0 & 1.6 & 0.91 \\ \hline
BR$_{\rm inv}$ (\%) & 0.23 & 0.22 & 0.61 & 0.024 \\ 
BR$_{\rm EXO}$ (\%) & 1.4 & 1.4 & 2.4 & 1.0 \\ \hline
\end{tabular} 
\end{table}
}

\begin{figure}[htbp]
\centering
\includegraphics[width=0.60\textwidth]{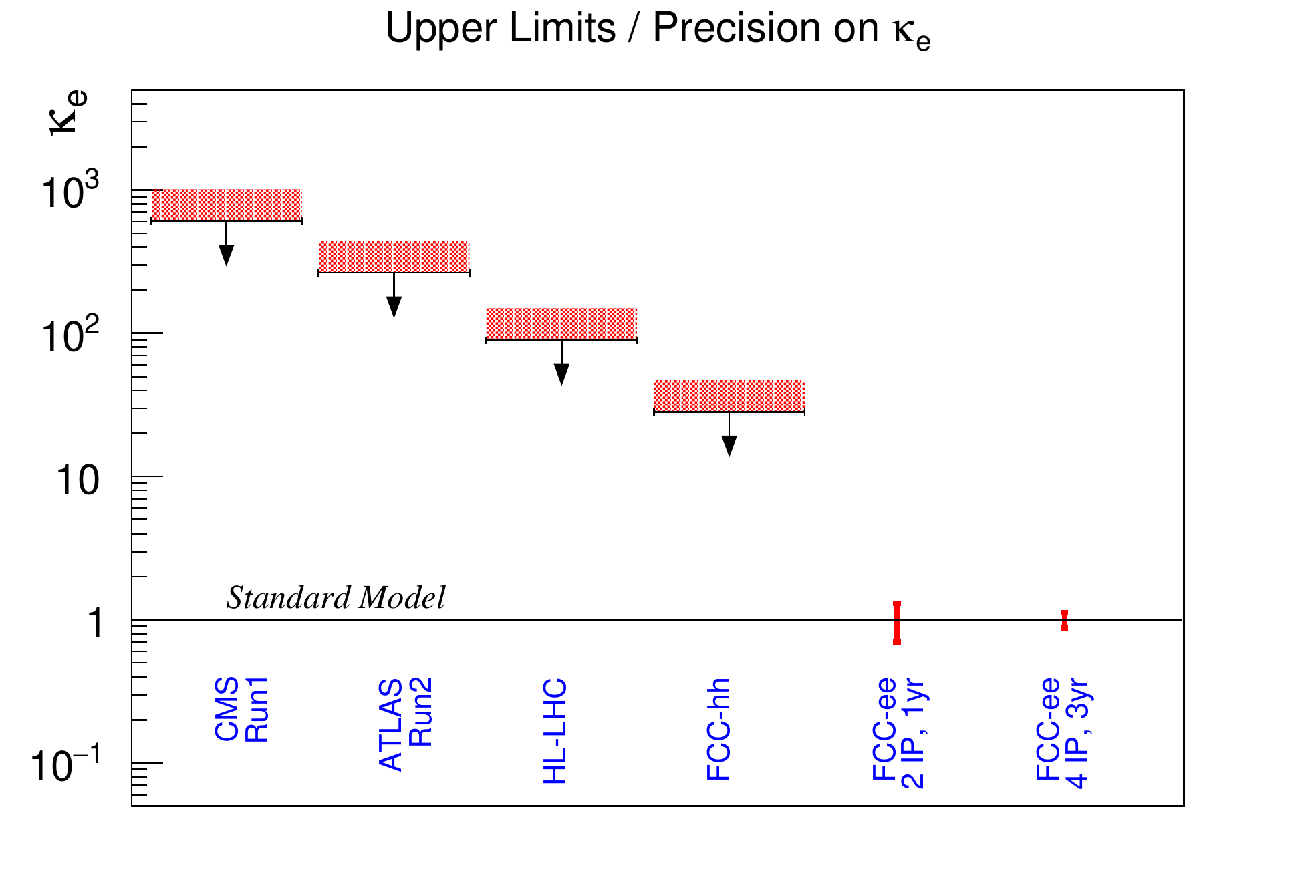}
\caption{\label{fig:Hee} \small Current upper limits on the Higgs boson coupling modifier to electrons, $\kappa_{\rm e}$, from CMS~\cite{Khachatryan:2014aep} and ATLAS~\cite{Aad:2019ojw}; projected $\kappa_{\rm e}$ upper limits at HL-LHC and FCC-hh; and projected $\kappa_{\rm e}$ precisions at FCC-ee in two different running configurations (one year with 2 IPs, or three years with 4 IPs).}
\end{figure}

Several remarks are in order. 
\begin{itemize}
    \item 
  Circular colliders operating at the maximum of the ZH cross section are very efficient Higgs factories, requiring $\sim 9$ ($5$)\,GJ per Higgs boson for FCC-ee with 2 (4) IPs at 240 GeV, vs $\sim 50$\,GJ for the proposed ILC run plan at 250 GeV. The measurement of the total ZH cross section at FCC-ee provides the most precise determination of the $g_{\rm HZZ}$ coupling. Linear colliders running at higher energies progressively obtain better measurement of $g_{\rm HWW}$ from the ${\rm H \nu_e \bar\nu_e}$ cross section. The synergistic combination of a circular and a linear colliders offers the best test of the SM relationship between the Higgs couplings to the W and Z boson, i.e., a test of the $\rm SU(2) $ custodial symmetry in the Higgs sector.
  
    \item 
 Proton-proton collisions are qualitatively and quantitatively more effective to study the Higgs boson thoroughly at high energy, once the $g_{\rm HZZ}$ coupling and the top-quark electroweak couplings (Section~\ref{sec:Top}) are determined in an absolute manner by (an) ${\rm e^+e^-}$ collider(s) operating at 240\,GeV and above 350\,GeV, respectively, and used as a standard candle in pp collisions. The FCC integrated plan yields precision consistently smaller than 1\% for 
 the couplings to gauge bosons and to fermions shown in Table~\ref{tab:EnergyUpgrades}, for the invisible and exotic branching fractions, and for the Higgs boson total width. With $5 \times 10^{10}$ Higgs bosons produced, FCC-hh also gives the most sensitive measurements of the rare decays such as $\mu\mu, \gamma\gamma, Z\gamma$, and of the invisible width. 
  
    \item 
  The Higgs self-coupling can be obtained by two different methods, as discussed in Ref.~\cite{deBlas:2019rxi}. The method with single Higgs production relies on the precise measurement of the ZH cross section (as shown in the right panel of Fig.~\ref{fig:Hsec}), and provides a robust determination from at least two sufficiently different energy points~\cite{DiMicco:2019ngk}. The need for these two energies provides a clear opportunity for a synergistic combination of a circular collider (operating up to 240\, GeV) and a linear collider (operating above 340\,GeV).
  
    \item 
  A more precise determination of the Higgs self-coupling can be obtained at higher energies from double Higgs production, at and above the ZHH cross-section maximum -- or at FCC-hh. This cannot be done with FCC-ee but is the realm of excellence of ILC at 1\,TeV or CLIC at 3\,TeV. An even better measurement is available at FCC-hh, which enjoys a wider phase space for double-Higgs production. For this important measurement, the different sources of systematic uncertainties in ${\rm e^+e^-}$ and pp collisions would also render their combination more robust than each individual result. 
    
    \item
  On the other side of the spectrum, the ability of FCC-ee to provide the highest luminosities at lower centre-of-mass energies opens the unique opportunity to measure the Higgs boson coupling to electrons~\cite{Ghosh:2015gpa,Dery:2017axi,Altmannshofer:2015qra} through the resonant production process ${\rm e^+e^- \to H}$ at $\sqrt{s} = 125$\,GeV~\cite{dEnterria:2017dac}. A $2\sigma$ excess (with respect to a situation in which the Higgs boson would not couple to electrons) would be observed at FCC-ee after a year with two interaction points, and a precision of $\pm 15\%$ on the Higgs boson SM coupling to the electron can be observed after three years with four interaction points. Because of the need of extremely high luminosity, combined with monochromatisation and continuous ppm centre-of-mass energy control, such a measurement is simply not possible with linear colliders. It is also out of reach of hadron colliders. A comparison with the hadron collider sensitivity is displayed in Fig.~\ref{fig:Hee}.
\end{itemize}

\subsection{Electroweak physics}
\label{sec:EW}

With the highest luminosities at the Z pole, the WW threshold, and the top-pair threshold, and with transverse polarisation to precisely calibrate the beam energies (and therefore to measure the Z mass, the Z width, or the W mass) with a precision of 100\,keV or thereabout, precision electroweak measurements are the realm of FCC-ee. The comprehensive set of precision electroweak observables that are accessible and measurable at FCC-ee provides a sensitivity to new physics with mass scale up to 70\,TeV, and may pave the way for FCC-hh direct searches. 

The tremendous impact of the Z pole data in the search for physics beyond the SM has been acknowledged by the recent ``revival'' of the GigaZ option~\cite{Yokoya:2019rhx,Fujii:2019zll} for ILC. With this option, and with longitudinal polarisation of both electrons and positrons, ILC could indeed compete, with uncertainties a factor between three and ten larger than FCC-ee, on a limited set of electroweak observables:  the left-right asymmetries of neutral-current couplings for the electron (${\cal A}_{\rm e}$),  with the measurement of the total cross-section beam-polarisation asymmetry; and for the other fermions (${\cal A}_\mu$, ${\cal A}_\tau$, as well as ${\cal A}_{\rm b}$ and ${\cal A}_{\rm c}$) from their forward-backward polarised asymmetries.   

The availability of accurate beam-energy calibration, on the one hand, and the three-order-of-magnitude higher statistics, on the other, allows FCC-ee to achieve factor 20--100 improvements (with respect to current measurements) of  the Z mass and the Z width~\cite{Blondel:2019jmp}, and of the QED coupling constant $\alpha_{\rm QED}(m_{\rm Z}^2)$~\cite{Janot:2015gjr} measurements, and to push  the sensitivity to new physics much beyond what would be available with ILC alone.  

\begin{figure}[htbp]
\centering
\begin{minipage}[b]{0.495\textwidth}
\centering
\includegraphics[width=\textwidth]{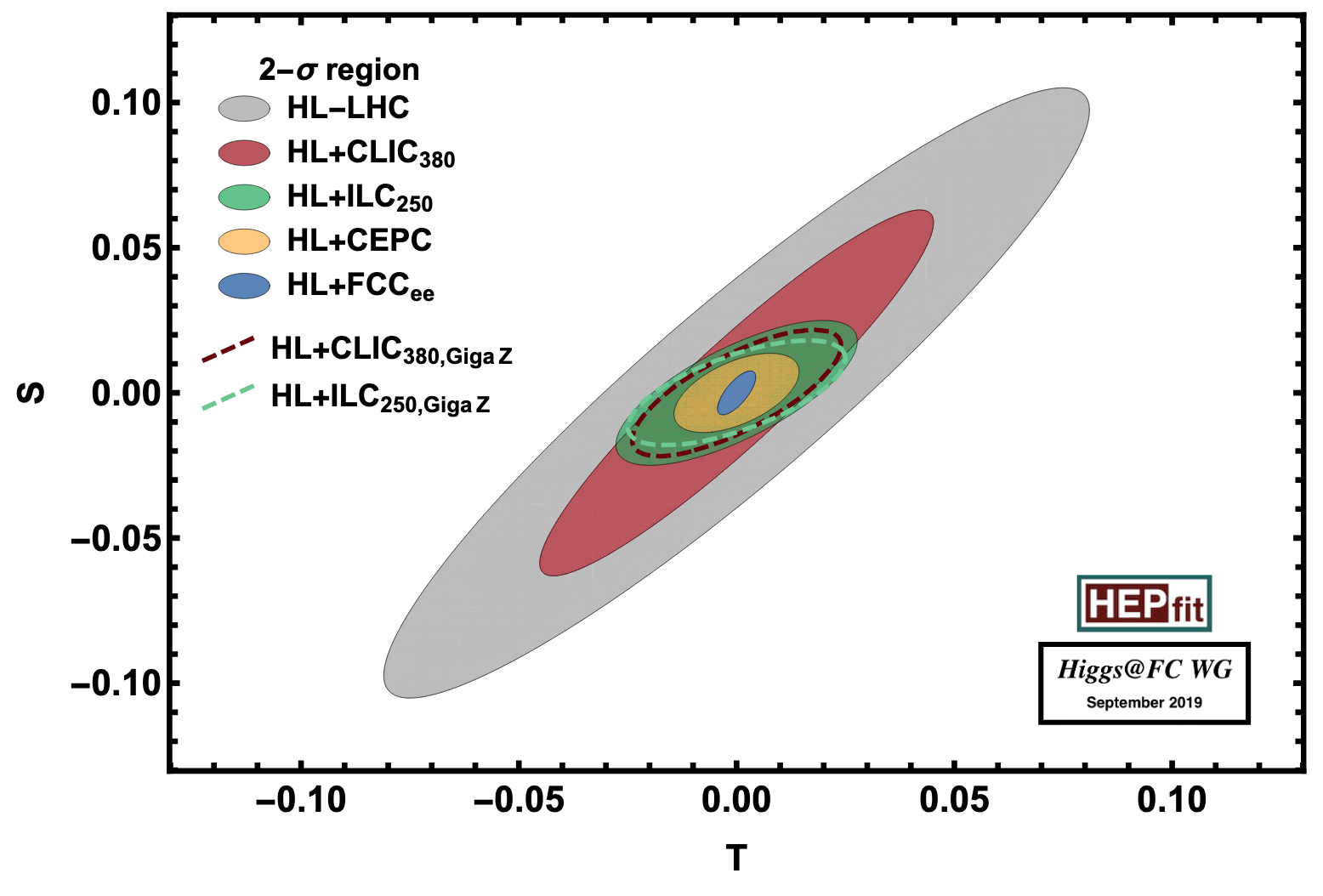}
\end{minipage}
\begin{minipage}[b]{0.482\textwidth}
\centering
\includegraphics[width=\textwidth]{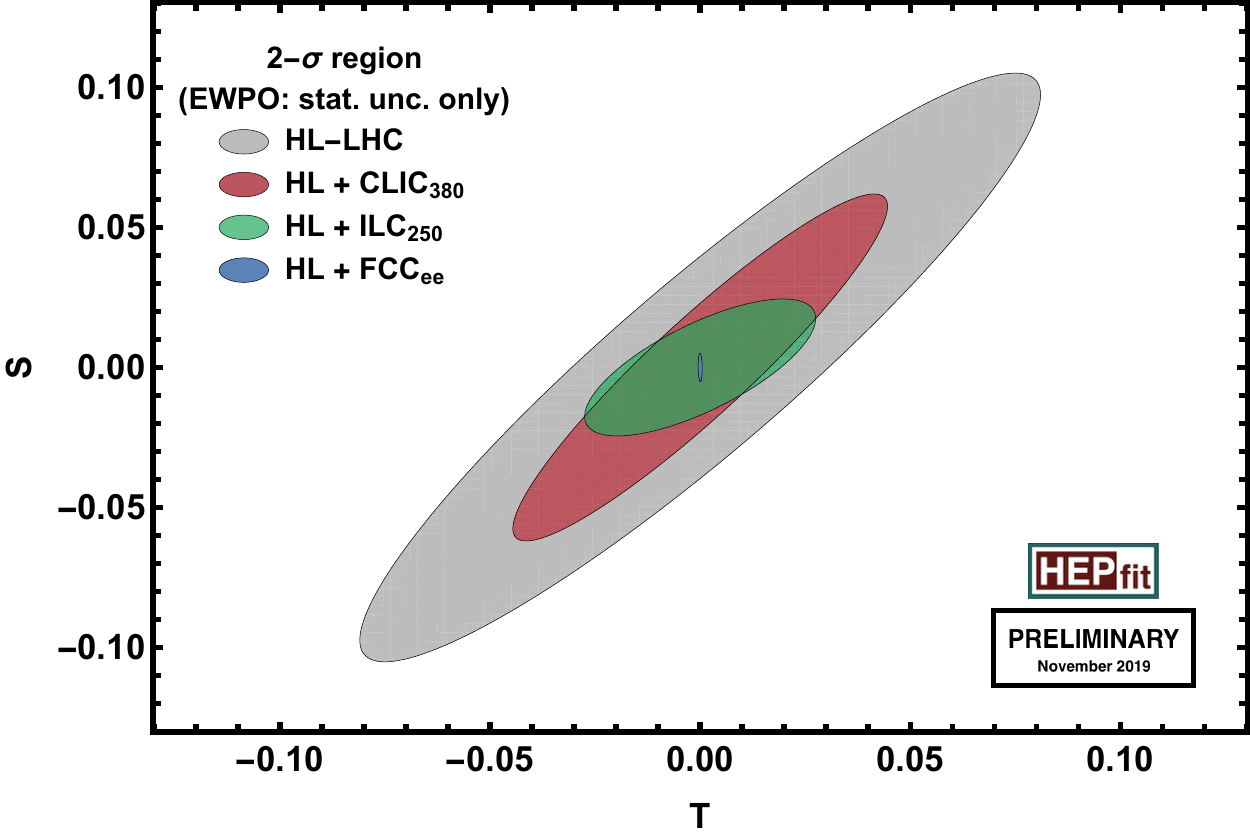}

\vspace{0.1cm}
\end{minipage}
\caption{\label{fig:EWPO} \small Expected uncertainty contour for the S and T parameters for various colliders in their first energy stage.  For ILC and CLIC, the projections are shown with and without dedicated running at the Z pole, with the current (somewhat arbitrary) estimate of future experimental and theoretical systematic uncertainty (left, from Ref.~\cite{Heinemann:2691414}); and with only statistical and parametric uncertainties (right, from J. de Blas).}
\end{figure}

This is illustrated in Fig.~\ref{fig:EWPO}, which shows the uncertainty on the $S$ and $T$ parameters (used to parameterise the isospin-conserving and isospin-breaking virtual effects in the Z and W propagators) from a global fit of the electroweak precision observables at the different ${\rm e^+e^-}$ colliders. In the left panel, with all experimental and theoretical systematic uncertainties included (as currently conservatively  evaluated), the ILC curves show that the higher-energy measurements (green ellipse) would not be improved with a dedicated run at the Z pole (dashed green ellipse). The FCC-ee measurements at the Z pole, at the WW threshold, and at the ${\rm t\bar t}$ threshold, on the other hand, allow the parametric uncertainties to be reduced to a minimum, as shown in the right panel, where only the statistical and parametric uncertainties are included. The ILC high-energy data would thus be optimally exploited only with the input from FCC-ee data.   

The right panel of Fig.~\ref{fig:EWPO} also shows the tremendous stand-alone true potential of FCC-ee, should future experimental and theoretical systematic uncertainties match the available statistics. The experiment and theory communities are set to devise proper detector specifications and complementary methods to reach these ambitious goals in the next 20 years.   

By their ability to reach higher energies, the linear colliders can access high-energy di-fermion and di-boson processes. This possibility provides sensitivity to new vector bosons and other high-mass objects via the so-called contact terms. The various sensitivities are compared in the Physics Briefing Book~\cite{Heinemann:2691414}, pp.\,114--118. In this domain, the high-energy $\rm e^+ e^- $ colliders can compete with either FCC-hh or a combination of FCC-hh with the precision measurements of FCC-ee, when expressed in terms of specific operators. It should be born in mind, however, that although in some cases they show similar sensitivities, the tests performed at high energies and those performed at the electroweak scale are  different in essence and sensitive to different effects, thus fundamentally complementary.

\subsection{Flavour Physics and QCD}
\label{sec:Flavour}

Progress in flavour physics require the huge data sample expected at the Z pole with FCC-ee, and therefore exemplify the need of a circular collider for a comprehensive set of measurements at the intensity and energy frontier.  

A total of about $10^{12}$ $\rm b \bar b$ pairs, available with a sample of $5 \times 10^{12}$ Z decays promised by FCC-ee, will challenge the precisions of CKM matrix element measurements expected from LHCb and Belle2, and push forward the search for unobserved phenomena such as CP-symmetry breaking in the mixing of beautiful neutral mesons~\cite{Benedikt:2651299}. 

In parallel, searches for rare decays make FCC-ee a discovery machine. Lepton-flavour-violating (LFV) Z decays, rare and LFV $\tau$ decays, searches for heavy neutral leptons and rare b-hadron decays have all been explored in Ref.~\cite{Benedikt:2651299} as benchmark or flagship searches, illustrative of the unique potential of a high-luminosity Z factory. The precision of the measurements relies on the vertexing capabilities of the experiments to take benefit of the boosted topologies at the Z energy, but most are limited in precision by the statistical size of the sample.  

A minimum of $5 \times 10^{12}$  Z decays has been shown to be necessary to make, for example, a comprehensive study of the rare electroweak penguin transitions $\rm b \to s \tau^+ \tau^-$~\cite{Kamenik:2017ghi}. For example, about 1000 events with a reconstructed $\rm \overline{B}^0 \to K^{\ast 0} \tau^+\tau^-$ are expected in such a sample. Should the current ``flavour anomalies"~\cite{Graverini:2018riw} persist, the study of b-hadron decays involving $\tau$'s in the final state is required to sort out possible BSM scenarios. If these flavour anomalies do not survive future LHCb and Belle2 future scrutiny, the study of Z couplings to third-generation quarks and leptons still constitute an excellent opportunity to unravel BSM physics. 

Tau physics at the Z pole provides, still today, the most powerful tests of lepton universality and of the PMNS matrix unitarity. The increase of statistics by five orders of magnitude with respect to LEP should allow much improvement of these constraints, by combining measurements of the tau lifetime, the tau mass, and the leptonic branching ratios. These measurements provide important constraints on e.g. light-heavy neutrino mixing. Similar improvements will result from the better precision of the  invisible Z decay width measurement~\cite{Benedikt:2651299, Benedikt:2653673, Abada:2019lih}. 

Finally, the $3.5 \times 10^{12}$ hadronic Z decay expected at FCC-ee also provide precious input for comprehensive QCD studies at the Z pole~\cite{Jorgen}. In particular, the determination of strong coupling constant $\alpha_{\rm S}(m_{\rm Z}^2)$, from the ratio $R_\ell$ of the Z hadronic width to the Z leptonic width, greatly benefits from the available statistics. An experimental uncertainty on $\alpha_{\rm S}(m_{\rm Z}^2)$ of $0.00015$ can be contemplated, with further improvement being actively planned for the next round of detector studies. A similar figure can possibly be obtained from tau decays, or from the measurements of the hadronic and leptonic decay branching ratios of the W boson~\cite{dEnterria:2016rbf}, copiously produced with FCC-ee operating at larger centre-of-mass energies. 

The energy evolution of event shapes and fragmentation functions also provide powerful tests of QCD in $\rm e^+ e^-$ collisions. The wide range of energies covered by FCC-ee, which can start at centre-of-mass energies as low as 30 GeV, in combination with high-energy domain of ILC provide a remarkable lever arm for such studies.  

\subsection{Top physics}
\label{sec:Top}

Most of what is known of the top quark today comes from hadron colliders. In ${\rm e^+e^-}$ collisions, the region of the top-pair threshold, and immediately above, is covered by both FCC-ee and ILC, with similar luminosity and energy efficiency (Fig.~\ref{fig:luminosities}). The main output of a run at the top-pair threshold is the precise measurement of the top-quark mass, which is an essential input to reduce parametric uncertainties in the electroweak precision measurements. The aforementioned precise $\alpha_{\rm S}(m_{\rm Z}^2)$ measurement at FCC-ee reduces the corresponding parametric uncertainty of the top mass measurement at the top-pair threshold, otherwise dominant e.g. at ILC alone. 

Running above the top-pair threshold, $\sqrt{s} = 365$\,GeV for FCC-ee and 500\,GeV for ILC\footnote{380\,GeV was found by the CLIC team to be optimal for the measurement of the top electroweak couplings.}, enables in addition the precise measurement of the top-quark couplings to the Z boson and the photon. This measurement is essential for the precise determination of the top Yukawa coupling at FCC-hh, through the measurement of the ${\rm t\bar t H}$ to ${\rm t\bar t Z}$ cross-section ratio. 

\begin{figure}[htbp]
\centering
\begin{minipage}[b]{0.54\textwidth}
\centering
\includegraphics[width=\textwidth]{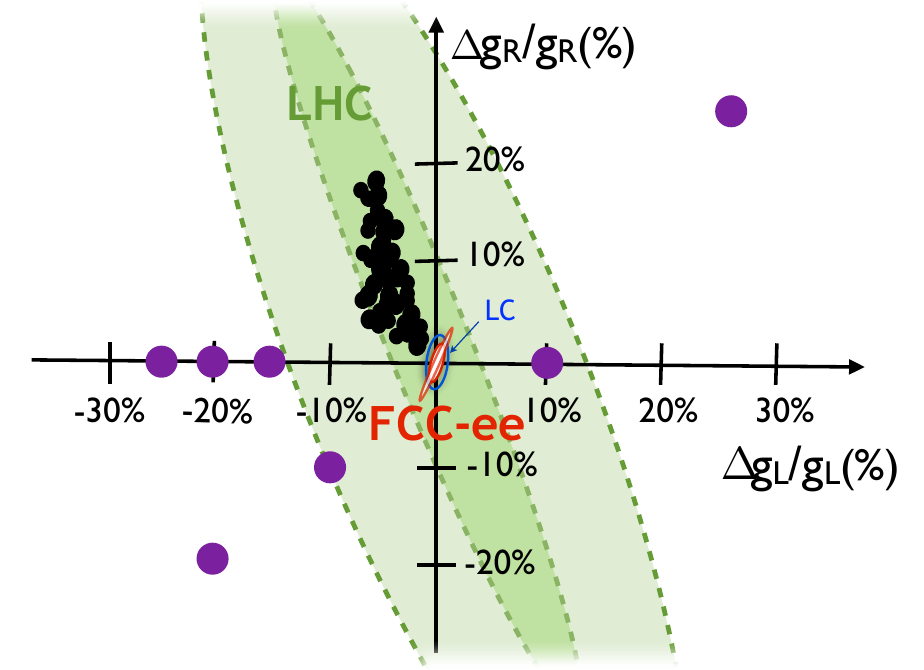}
\end{minipage}
\begin{minipage}[b]{0.45\textwidth}
\centering
\includegraphics[width=\textwidth]{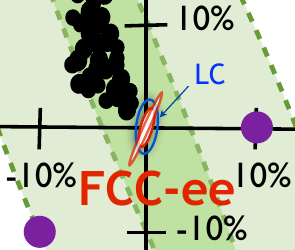}

\vspace{0.1cm}
\end{minipage}
\caption{\label{fig:topEW} \small (Adapted from Ref.~\cite{Barducci:2015aoa} and taken from Ref.~\cite{Janot:2015mqv}) Expected relative precision on the ${\rm Zt_Lt_L}$ and ${\rm Zt_Rt_R}$ couplings at LHC (lighter green), HL-LHC (darker  green), ILC at 500\,GeV (blue) and FCC-ee (orange, red). The right panel shows a zoom of the left panel in the $\pm 10\%$ window. Typical deviations from the standard model in various new physics models are represented by the purple dots. The black dots indicate the deviations expected for different parameter choices of 4D  Composite Higgs Models, with mass scale up to $2$\,TeV. For FCC-ee, the orange ellipse is  obtained from an analysis of the lepton angular and energy distributions, while the smaller red ellipse is obtained if the angular and energy distributions of the b jets can be  exploited.}
\end{figure}

The original ILC method exploits the longitudinal beam polarisation~\cite{Baer:2013cma} to disentangle the couplings of the left-handed and right-handed top quark to the Z and the photon. For FCC-ee, it was shown that the final-state top-quark polarisation can be used effectively~\cite{Janot:2015yza}, to the same effect. Based on the latter study by FCC-ee, a more recent ILC study~\cite{Sato:2018ahh} also uses the final-state polarisation information, and observes a 10 to 40\% improvement with respect to using initial polarisation only. (This kind of ``sociological'' complementarity is addressed in more detail in Section~\ref{sec:sociological}.) These two complementary approaches lead to similar sensitivity but with different correlation coefficients, as shown in Fig.~\ref{fig:topEW}. A combination of ILC and FCC-ee data would therefore be beneficial to the precise measurements of the top Yukawa coupling. 

The access to centre-of-mass energies of 500\,GeV and above opens the additional possibility of determining simultaneously the ten form factors that parameterise the ${\rm t\bar t Z}$ and ${\rm t\bar t} \gamma$ couplings~\cite{Khiem:2015ofa,Durieux:2018tev,Durieux:2019rbz}, including the four CP-violating form factors, while the FCC-ee run at 365\,GeV can determine only four CP-conserving parameters simultaneously and one linear combination of CP-violating form factors. The present status of circular collider studies does not allow very firm conclusions to be reached on the determination of (either standard or non-standard) couplings of the top quark. An impressive analysis of this interesting topic using full event-by-event information has been performed in the framework of ten EFT-based free parameters~\cite{Durieux:2018tev} to understand the relative merits of {\it (i)} beam energy; {\it (ii)} beam longitudinal polarization; and {\it (iii)} final-state lepton and quark flavours and charges. 

Finally, as shown in Table~\ref{tab:EnergyUpgrades}, an independent measurement of the top Yukawa coupling, complementary to the HL-LHC (FCC-hh) measurement at the 3\% (1\%) precision level, can also be obtained with ${\rm e^+ e^-}$ data at 500\,GeV (with $\pm 6.9\%$ standalone precision) or at 1\,TeV (with $\pm 1.6\%$ standalone precision).  

\subsection{Physics beyond the Standard Model}
\label{sec:BSM}

Particle physics has reached an important moment of its history. On the one hand, the Standard Model is a complete and consistent theory that describes all observed collider phenomena, and that can be extrapolated to the Planck scale without the need of new physics. On the other hand, observations that cannot be explained by the standard model (e.g., the existence of dark matter, the baryon asymmetry of the universe, the smallness of neutrino masses) require particle physics explanations. For the first time since Fermi theory, however, there is no clear mass scale for this new physics, and the strength of its couplings to Standard Model particles is just unknown. Without specific target, the search for new physics must therefore be as broad and as diverse as possible, with 
\begin{enumerate}
    \item the measurements of small deviations from precise predictions (similar to the top-quark and Higgs-boson  mass predictions from the measurements of radiative corrections);
    \item the observation of new phenomena (similar to, e.g.,  neutral currents, neutrino oscillations or CP violation);
    \item the direct search and the observation of new particles, with {\em a priori} any mass and any coupling to the standard sector.
    \end{enumerate}

\begin{figure}[htbp]
\centering
\includegraphics[width=0.85\textwidth]{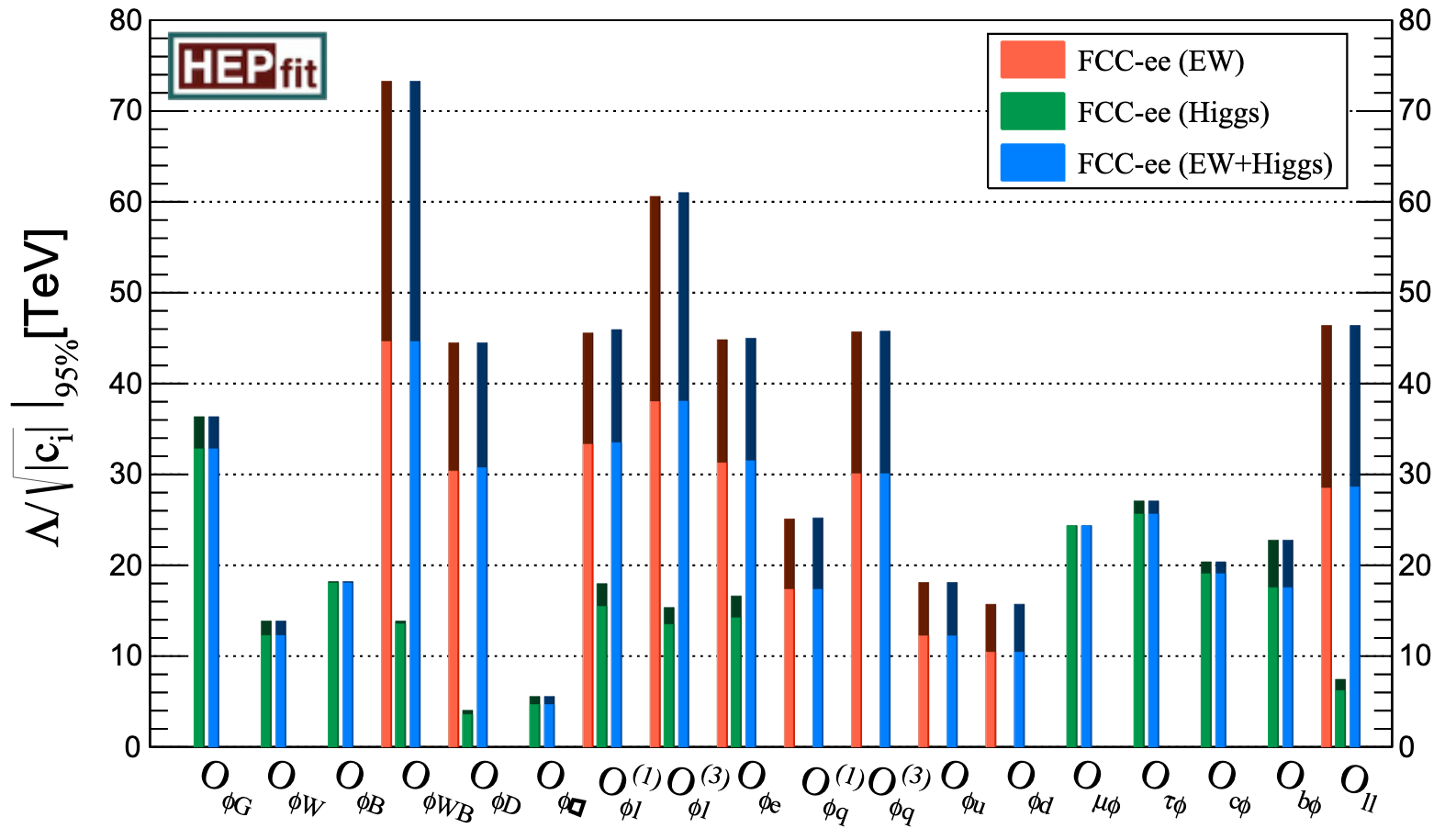}
\caption{\label{fig:Jorge} \small Electroweak (red) and Higgs (green) constraints from FCC-ee, and their combination (blue) in a global EFT fit. The constraints are presented as the 95\% probability bounds on the interaction scale, $\Lambda/\sqrt{c_i}$,  associated  to  each  EFT operator. Darker  shades  of  each  colour indicate the results when neglecting all SM theory uncertainties}
\end{figure}

An across-the-board exploration of these three aspects can be performed effectively with the association of FCC and ILC operating at different centre-of-mass energies. As discussed in Sections~\ref{sec:Higgs} and~\ref{sec:EW}, measurements of small deviations of Higgs properties or of precision electroweak observables benefit from the ILC/FCC-ee complementarity. The sensitivity of these measurements to heavy new physics can be quantified with a global EFT fit, as shown in Fig.~\ref{fig:Jorge}. This figure, made for FCC-ee only, highlights {\it (i)} the complementarity of Higgs and electroweak measurements; and {\it (ii)} the need of more statistics for the (statistics-limited) Higgs measurements, thus confirming the potential benefits of combining optimally the ILC and FCC-ee data. 

The effect of a heavy ${\rm Z}^\prime$ gauge boson on the ${\rm e^+e^-} \to \mu^+\mu^-$ process provides an illustrative example of this complementarity. The 100 times more precise measurements (with respect to LEP) of cross sections and asymmetries by FCC-ee at and around the Z pole would be sensitive to such a new object by ${\rm Z/Z}^\prime$ mixing or interference. Measurements with higher-energy ${\rm e^+e^-}$ collisions would display increasing deviation of the dilepton (or other diquark or diboson) cross sections with respect to the standard model prediction, when increasing the centre-of-mass energy. The combination of the two effects would be a tell-tale signal and would then allow the mass and the couplings of this new $Z^\prime$, as well as other model parameters, to be determined. The measurement of the top electroweak couplings would then complete the landscape and help identifying the underlying theory. (See for example Ref.~\cite{Benedikt:2651299} for a specific Higgs composite model.)  

\begin{figure}[!ht]
\centering
\includegraphics[width=0.9\textwidth]{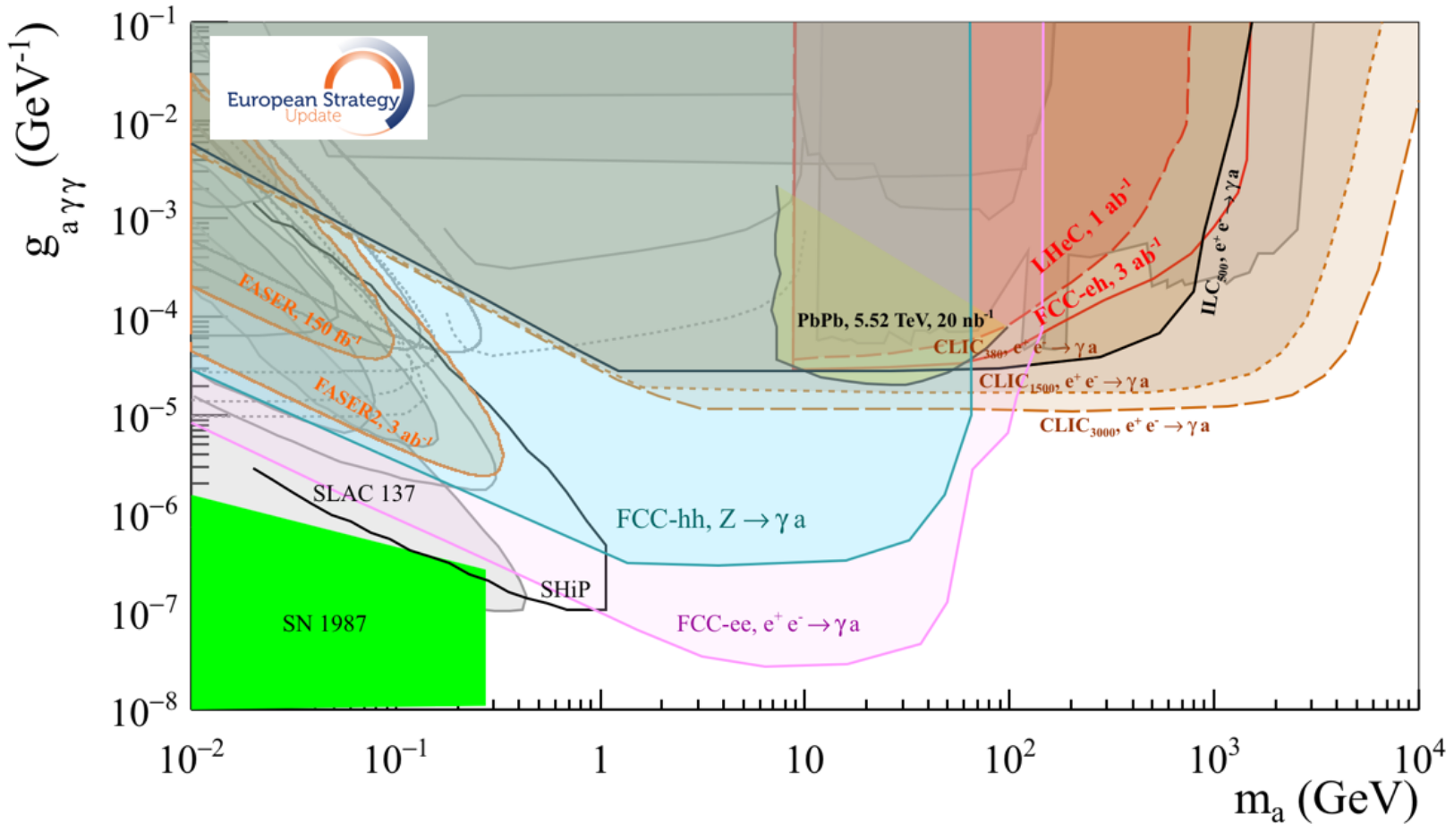}
\caption{\label{fig:ALPS} \small Expected sensitivity to Axion-like particles in various future facilities. The reach of FCC-ee is at very small couplings in Z decays, while the reach of linear colliders is at higher masses for somewhat larger couplings. From Ref.~\cite{Heinemann:2691414}  }
\end{figure}

Direct observation of new particles might require very diverse searches and colliders to cover many orders of magnitude of coupling strength and mass scales. Feebly interacting particles such as  axion-like particles, or dark photons, with masses smaller  than the Z mass, may have the best odds of being found at the intensity frontier, i.e., with FCC-ee running at the Z pole. High-energy machines can see them if their couplings are larger. The case of the axion-like particles is illustrated in Fig.\ref{fig:ALPS}, and demonstrates the typical complementarity between the Z factory FCC-ee and a high-energy linear $\rm e^+ e^-$ collider. 

\begin{figure}[!ht]
\centering
\includegraphics[width=0.9\textwidth]{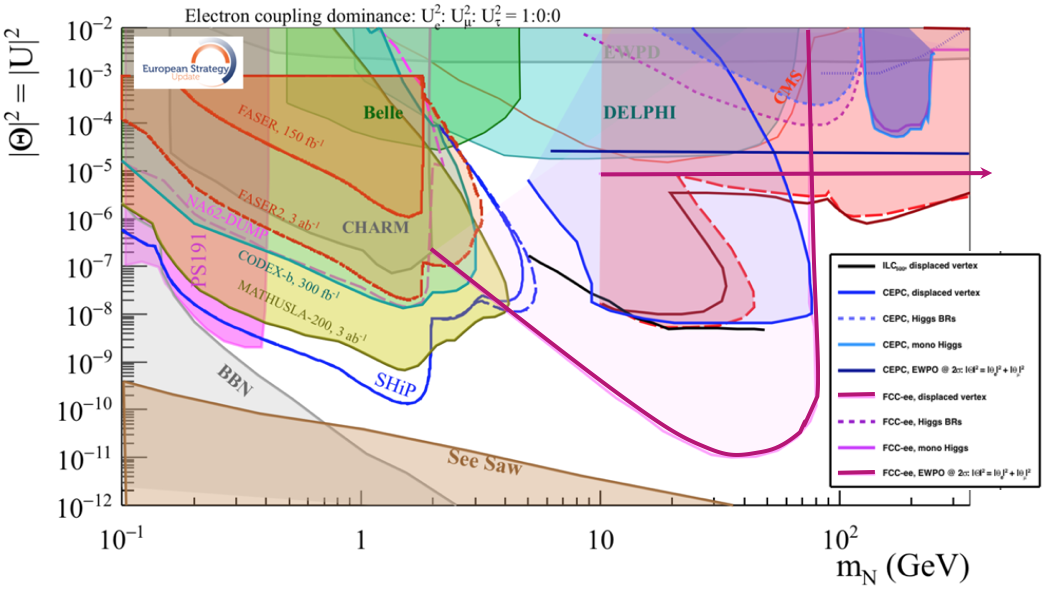}
\caption{\label{fig:RHnu} \small Expected sensitivity to Heavy-Neutral Leptons (a.k.a. Right Handed Neutrinos) in various future facilities. The reach of FCC-ee is for very small heavy-light mixing angle in Z decays, down to the see-saw limit; it is complemented up to very high masses (60 TeV or more) for heavy-light neutrino mixing larger than $10^{-5}$ by constraints from Electroweak and tau decay precision measurements. See~\cite{Heinemann:2691414}, Fig 8.19.}   
\end{figure}

Another well-motivated example of new physics is provided by neutrinos. Many neutrino mass models naturally predict the existence of heavy neutrino states, called Heavy Neutral Leptons (HNL, mostly of right-handed chirality or ``sterile'') which mix with the known light, active neutrinos with a typical mixing angle  $|\theta_{\rm \nu N}|^2 \propto  m_{\nu}/m_{\rm N}  $. Since both light and heavy neutrino masses are unknown, a rather large range of mixing angles should be explored.   These scenarios  have several possible consequences: {\it (i)} the direct observation of a long-lived HNL in Z, W, and Higgs decays and in tau, b- or c-hadron semi-leptonic decays, both mass and mixing sensitive; {\it (ii)} the mixing of the light neutrinos with heavier states, which leads to a violation of the SM relations in EWPOs; the corresponding sensitivity only depends on the mixing angle, and extends to very high masses; {\it (iii)} the violation of lepton universality in $\tau$, b or c-hadron decays at the Z factory; {\it (iv)} a deficit in the Z invisible decay width; and {\it (v)} a lepton-number violation can also result from Heavy-Neutral-Lepton production or exchange in high-energy processes at a hadron collider or a high-energy ${\rm e^- e^-}$ collider. The most sensitive tests {\it (i)} and {\it (ii)} for masses above $m_{\rm N} \geq 10$\,GeV are performed at FCC-ee, as shown on Fig~\ref{fig:RHnu}.

\begin{figure}[!htb]
\centering
\includegraphics[width=0.40\textwidth]{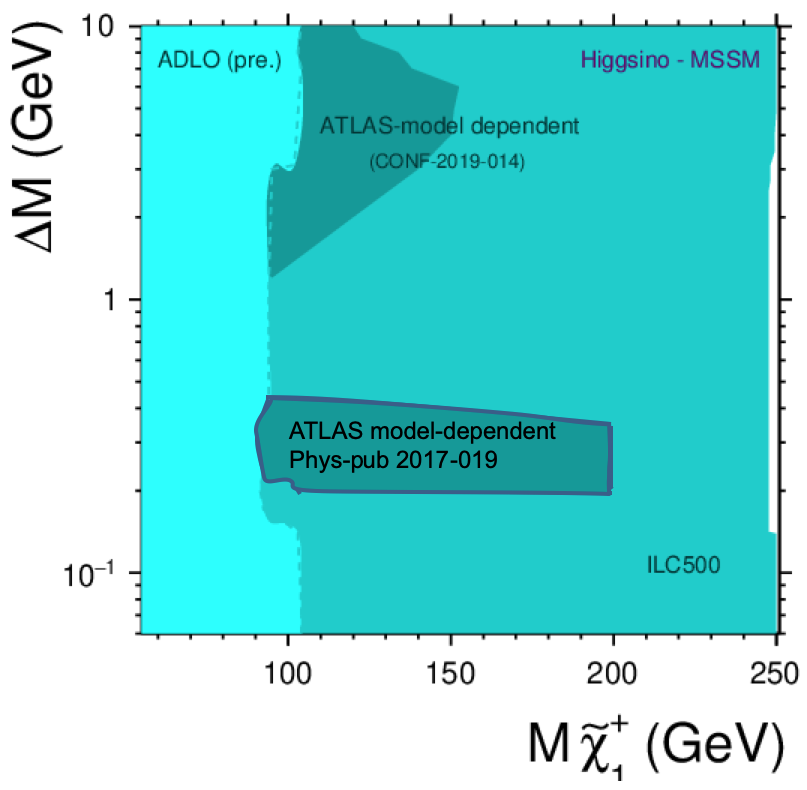}
\includegraphics[width=0.49\textwidth]{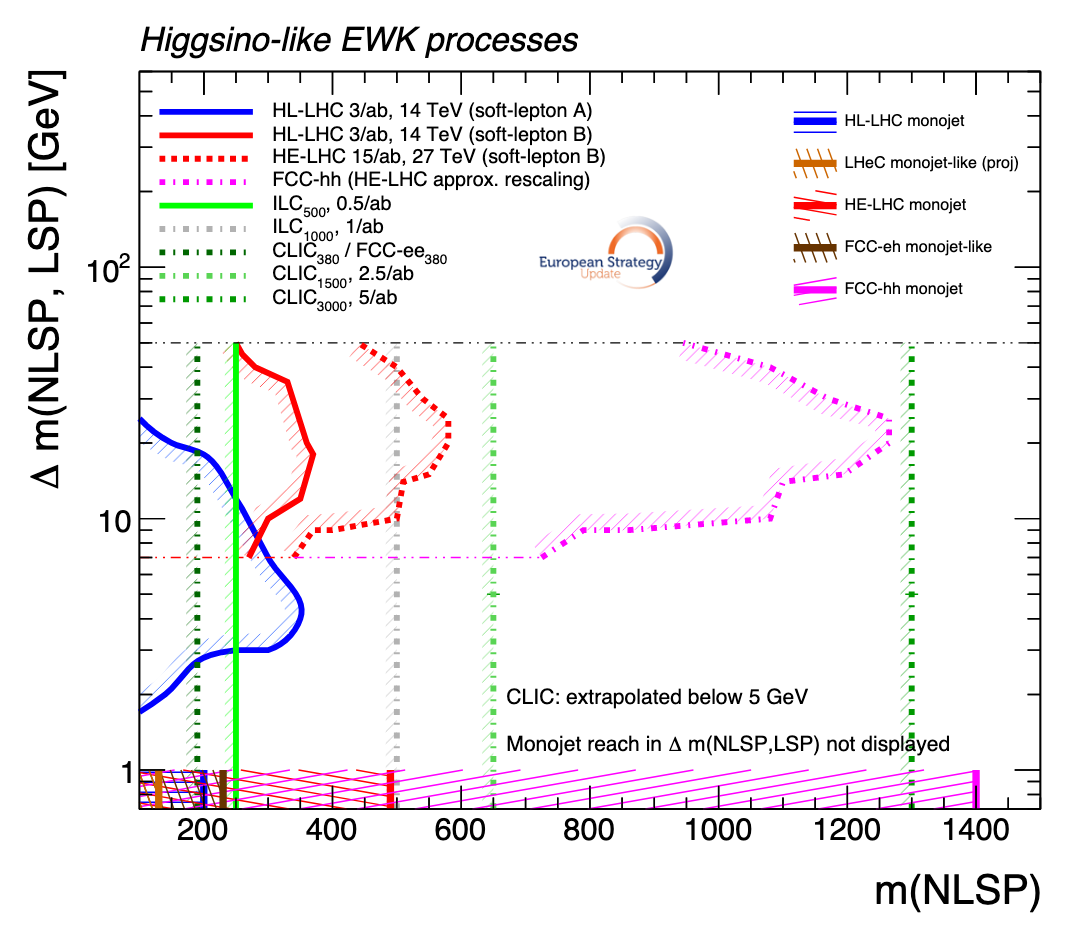}
\caption{\label{fig:Higgsino} \small Left: Higgsino sensitivity at ILC$_{500}$. From J. List, KET-Jahrestreffen, 14 November 2019, T. Nunez, LCWS2019. Right: {Fig. 8.10 from the Physics Briefing Book~\cite{Heinemann:2691414}. Exclusion reach for Higgsino-like charginos and next-to-lightest neutralinos with equal mass m(NLSP), as a function of the mass difference $\Delta m$ between NLSP and LSP. Exclusion reaches using monojet searches at pp and ep colliders are only displayed up to  1 GeV, but are expected to extend up to $\Delta m$ of several tens of GeV$^4$ in both panels.}}
\end{figure}

Weakly interacting particles such as charginos with masses of a few hundred GeV, might have escaped the LHC searches -- and might continue to escape the HL-LHC and FCC-hh searches -- in case of compressed supersymmetric mass spectra.\footnote{It is argued that Fig.~8.10 of the Physics Briefing Book~\cite{Heinemann:2691414}, displayed for reference in the right panel of Fig.~\ref{fig:Higgsino}, is misleading in this respect: small prints on this figure say ``Monojet reach in $\Delta m$ is not displayed'', but the monojet reach is expected to cover the whole $\Delta M$ region as well. See for example Fig.~6 of Ref.~\cite{Sirunyan:2019zfq} obtained with $36\,{\rm fb}^{-1}$ of CMS data at LHC, which covers the wino scenario up to 100\,GeV for $\Delta M = 1$\,GeV and above, and which extrapolates with FCC-hh to around 1\,TeV for $\Delta M = 1$\,GeV and above in the Higgsino scenario~\cite{Filip}. The constraints from the monojet search are expected to be even more stringent. If confirmed, such a statement would somewhat affect the linear-collider complementarity with FCC-INT.} Such particles, however, could affect precision electroweak measurements and flavour physics, so that the Z pole run of FCC-ee can bring considerable constraints on the parameter space. Linear colliders with centre-of-mass energies between a few hundred GeV to a few TeV are likely to offer the best shot at finding these particles, with masses below $\sqrt{s}/2$, as demonstrated in the left panel of Fig.~\ref{fig:Higgsino}. Finally, coloured particles, Higgs bosons, or weakly interacting particles with masses of 5 to 50\,TeV require the largest parton-parton centre-of-mass energy offered by hadron colliders. A complete coverage requires all of them.

\subsection{Summary}
\label{sec:summary}
The scientific complementarity of FCC-ee and ILC is summarized in Table~\ref{tab:complementarity}.

The two projects share access to the ZH  cross-section maximum (240-250 GeV), so results obtained in this region are not included. The table reveals, however, very significant domains of physics where circular and linear $\rm e^+ e^-$  colliders at the electroweak scale are clearly different. These domains offer considerable opportunities of discovery. For FCC-ee,  precision measurements and search for rare processes at the Z are unique, and so is the possibility to access the Higgs-electron coupling; this is a domain where FCC-ee is highly complementary with FCC-hh. For ILC, the specificity lies in the ability of accessing energies where the search gaps of the hadron colliders can be neatly covered; this capacity is, by construction, complementary with FCC-hh as well. 
These areas of complementarity between FCC-ee and ILC are therefore also complementary with the capacities of the hadron collider program. 

\vfill\eject
{\null}
{\setlength{\tabcolsep}{6pt} 
\renewcommand{\arraystretch}{1.1} 
\begin{table}[!htbp]
\fontsize{9}{10.8}\selectfont
\centering
\caption{\label{tab:complementarity}\small Summary of complementary qualities of the proposed circular and linear colliders FCC-ee and ILC. Notes: $^1$ single-parameter sensitivity, full program; $^2$ multi-parameter sensitivity up to 365/500\,GeV; LFUV: Lepton Flavour Universality Violation; LNV: Lepton Number Violation.}
\label{tab:eeomp}
 
\vspace{2mm}
\begin{tabular}{|l|c|c|}
\hline
Quality  & FCC-ee & ILC  
\\ 
\hline \hline
{\bf Energy Range (GeV)} & 88 to 240,  up 365  & (91) 240  up 500, 1000   \\
\hline \hline 
{\bf Interaction points} & 2--4 & 1   \\
\hline \hline
{\bf Luminosity} &  ${\propto E_{\rm beam}^{-3.5}\times {\rm Radius \times Power\times \#IP}} $ & ${\propto E_{\rm beam} \times {\rm Power}}$   \\

Main statistics & &  \\ 
 Z  & ${\rm 5.10^{12}~Z}$ & ${\rm 5.10^{9}~Z}$  \\
 WW & ${\rm  3.10^8~WW}$ & ${\rm  10^7~WW}$   \\ 
 HZ &  ${ \rm 10^6~H }$ & ${ \rm 4.10^5~H }$   \\ 
${\rm t\bar{t} }$ and above & ${\rm 10^6~t\bar{t}}$ at 365\, GeV & ${\rm 3.10^6~t\bar{t} }$ at 500\,GeV \\
\hline \hline 
{\bf Beam Polarisation} &  Transverse   & Longitudinal     \\
For               & ${\rm e^+ and~ e^-}$ & $\rm e^- (\pm 80\%)$, $\rm e^+ (\pm 30\%)$ \\
Beam Energies    & up to WW threshold & all energies  \\
Use & $\sqrt{s}$  ppm calibration  & helicity cross-sections \\ 
\hline \hline 
{\bf Monochromatisation} &  $\rm \sigma_{\sqrt{s}}= 4-10$~MeV   
& no  \\
Use                     & $s$-channel H production& \\ 
            \hline \hline 
 {\bf Higgs Physics}  &  & \\
 Hee Coupling  & SM ($m_{\rm e}$) $\pm$ 15-50\%    & --  \\
 HHH Coupling:  &   &   \\
 ~~ 
 from $\sigma({\rm e^+e^- \to ZH})$   & $\pm 14^1-33^2$\%  & $\pm 25^1-38^2$\%    \\
 ~~ from HH production  & --  &  $\pm$ 27\% (500\,GeV), $\pm$ 10\% (1\,TeV) \\
 \hline \hline
 &  $m_{\rm Z}, \Gamma_{\rm Z}$, $m_{\rm W}$ (100, 25, 600\,keV)  & High-energy polarised   \\
 {\bf Electroweak }  & ${\rm \sin^2\theta^{eff}_{W} (3.10^{-6})}$ ${\rm \Delta{\alpha_{QED}}  (3.10^{-5})}$  & Cross sections and asymmetries  \\
 & LFUV  $g_{\rm A}~(10^{-5}),~ g_{\rm V}~(10^{-5})$& for leptons, quarks and bosons \\
  & EFT operators up to 70 TeV  & contact interactions up to 100 TeV     \\
 \hline \hline
 &$e / \mu / \tau$  LNV $10^{-10}$ &\\
  {\bf Flavour Physics} & LFUV   $< 10^{-5}$  &   \\
  & b and c hadrons  properties&  \\
 & rare decays and CPV  & \\
 \hline \hline
 & 30-365 GeV jet systems   & 240-1000 GeV jet systems  \\
 {\bf QCD } &  hadronisation & hadronisation  \\
 & $\alpha_{\rm s}$ in Z,W,$\tau$ ($10^{-4}$) &  \\
 \hline \hline
 & in Z decays: &  up to 500 GeV pair production\\
 {\bf New particle search} &  Feebly coupled particles  &  searches in gaps left by  \\
 & RH neutrinos, ALPs etc.  & hadron collider \\ 
\hline \hline
\end{tabular} 
\end{table}
} 
\noindent The main points of the table can be further summarized as function of the energy range:
\begin{enumerate}
    \item {\bf At energies between 30 and 240\,GeV:  FCC-ee}
    \begin{itemize}
    \itemsep-0.1em 
    \item[--] High luminosity for Z, Hee, WW, HZ; 
    \item[--] Exquisite energy calibration at the Z, Hee, WW; 
    \item[--] Monochromatisation at $\rm \sqrt{s}= m_H$; 
    \item[--] Z and W factory with $5.10^{12}$Z and  $3.10^8$ WW, enabling electroweak measurements, ${\rm \alpha_{QED}}$ and ${\rm \alpha_{QCD}}$ determination, flavour (b, c, $\tau$) studies, QCD physics, searches for SM symmetry violations and feebly coupled particles: RHnu, ALPS, etc;
    \item[--] $s$-channel Higgs production: $g_{\rm Hee}$ coupling  
    \end{itemize}  

\item {\bf At energies between 240 and 380\,GeV: Both FCC-ee and ILC} 
    \begin{itemize}
    \itemsep-0.1em 
    \item[--] Higgs main couplings determined from copious decay modes in a model-independent way, with clear advantage for FCC-ee at 240\,GeV due to 5-to-10 times higher luminosity;
     \item[--] Higgs self-coupling inferred from ZH cross section energy dependence, also benefiting from larger luminosities (up to $4\sigma$ significance for FCC-ee);  
     \item[--] Measurements of $m_{\rm top}$ to better than $20$\,MeV. Determination of top neutral-current couplings, with some specificity for ILC because of beam polarization.  
    \end{itemize}     
    
\item {\bf At energies above 380\,GeV: ILC}
    \begin{itemize}
    \itemsep-0.1em 
    \item[--] Determination of Higgs self-coupling from double Higgs production;
    \item[--] Searches for new particles in the gaps left by hadron colliders. May be essential if a new particle is discovered at the hadron collider in ILC energy range;
    \item[--] High energy EW processes: lepton, quarks and boson pairs, with polarised beams
    \end{itemize}     

\end{enumerate}    
\section{Financial complementarity}
\label{sec:financial}

The presence of two projects in the world would of course lead to a larger burden on the international community, with the collateral advantage of raising the overall profile of the worldwide high-energy frontier community to a level more appropriate to support the ultimate high-energy proton collider. It is nevertheless useful to consider some possible optimisations and savings resulting from the availability of two complementary programmes. It can be expected that the run plan can be optimised in several different ways, following the following observations:
\begin{enumerate}
    \item The GigaZ run of ILC, even with the longitudinal polarisation, is made redundant by the TeraZ run of FCC-ee. This run can therefore be avoided.  
    
  \item Some of the most critical results at 240-250 GeV, including the ZH total cross section, could be carried out at both facilities with different detectors and different experimental conditions.  Cross-checks can therefore be safely performed without having two ILC detectors in push-pull configuration. Alternatively this region could solely be covered by a circular $\rm e^+ e^-$ collider, where is it more efficient, in order to concentrate the operation of a linear collider at the higher energies, where it is unique.

    \item The $\rm \bar{t}t$ run of FCC-ee is not considerably different from that of ILC. A concurrent running at both facilities just above the top-pair threshold would therefore allow a significantly better precision on the Higgs self-coupling with single Higgs production and on the top neutral-current couplings to be reached in the same time. Alternatively, the FCC-ee run at 365\,GeV could be replaced by a twice longer ILC run at a similar energy, in order to concentrate the FCC-ee operation at lower energies, where its capacity is unique.

\end{enumerate}

The resulting savings made by this coordinated distribution of tasks would allow a more efficient completion of the program, with the redistribution of run time to statistically limited measurements, such as the s-channel Higgs run at $\sqrt{s} = m_{\rm H}$ or the W mass measurement, at FCC-ee. Finally, the narrower domain of energies, that the detectors on both sides would have to cover, should lead to significant savings and to better optimisation. 

The monetary savings could altogether be non-negligible. The upgrade to the $\rm t \bar t$ threshold of FCC-ee has been estimated at 1.1\,BCHF. On the ILC side, {\it (i)} the additional re-arrangement of ILC for GigaZ; {\it (ii)} the extra work and investment required for positron polarisation at the lower  energies; and {\it (iii)} the savings on the detector push-pull; could add up to the best part of a billion USD. Further optimisation of the run plans, taking into account e.g. the actual integrated power required to produce a Higgs boson, or the best centre-of-mass energies required for the double Higgs measurements, can also be envisaged. 

\section{Sociological complementarity}
\label{sec:sociological} 

Should there be enough resources, the overlap in the centre-of-mass energies efficiently accessible at both FCC-ee and ILC, from 240 to 365\,GeV, would provide direct competition and cross motivation, and would give many opportunities of cross checks. This argument is less quantifiable and may therefore be more arguable than those presented in the previous section, but is interesting to expose here anyway for further discussion. 

Competition has consistently proven, in the past, to be a stimulus for excellence and breakthroughs. High-energy physics is no exception. The past five years of future-collider design studies bulge with resounding examples. Striking illustrations include {\it (i)} the seven-fold ILC luminosity increase at $\sqrt{s} = 250$\,GeV from $0.82$ to $5.4 \times 10^{34}\,{\rm cm^{-2}s^{-1}}$ (see footnote of Table~I in Ref.~\cite{Bambade:2019fyw} and Table 10.2 of the Physics Briefing Book~\cite{Heinemann:2691414}) since the ILC TDR publication; or {\it (ii)} the successive CEPC design evolutions (from 50 to 100 km circumference; from single to double ring; from a Higgs factory to a full electroweak factory with Z pole and WW threshold running; with a 100-fold increase of the luminosity at the Z pole in the past couple years; etc.). This progress might be attributed to competition with, e.g., FCC-ee. 

The same has been true for LEP and SLC, when operating at the Z pole at the end of the last millennium. Each collider was motivated by the other to deliver the highest luminosities, to develop the best detectors, to propose the best upgrades, to exploit specific assets of the circular and linear geometries (such as transverse and longitudinal beam polarisation), etc. Similarly, each of the four LEP collaborations was compelled to develop the brightest ideas in order to surpass the other three at the winter or summer conferences.  The opportunity of cross-checking and combining each other's results has also been a decisive asset altogether, in view of convincing the world of the robustness of the discovery that there was only three species of light neutrinos, or of the top-quark mass and Higgs-boson mass predictions. 

The situation will be very similar with contemporary construction and operation of FCC-ee and ILC. Who will observe the first Higgs boson in ${\rm e^+e^-}$ collisions? Who will first reach the design luminosity? Who will produce measurements with the best individual precision?  However anecdotal these sociological considerations may appear, they will unavoidably motivate the teams to give their best, go beyond what was initially intended, and produce in turn the best scientific return-on-investment by cross-checking and ultimately combining their results -- for what will be indeed the most precise measurements.  

From another angle, the ILC operation is expected to span over many decades, if the full programme up to 1\,TeV is to be completed. It would therefore cover the 10-years gap expected between the end of FCC-ee physics and the beginning of FCC-hh physics, ensuring continuous presence of data taking at the high-energy frontier. 

Last but not least, a situation where data can be collected in more than one experiment is preferable. A circular collider fills this preference internally, and in an efficient way, since luminosity grows almost linearly with the number of experiments. By providing redundant information for part of the  ILC program, it might also save the need for the somewhat inefficient push-pull layout at the linear collider. 

\section{Regional complementarity}
\label{sec:regional}
Finally, it will be appreciated that the opening of an era of $\rm e^+ e^- $ colliders would lead not only to the development of two very challenging accelerator concepts, but also to the creation of two novel infrastructures in the world. This would give promising scope for development of complementary programs with a long-term vision of high energy frontier colliders.

\section{A complementary FCC-ee / ILC operation plan?}
\label{sec:scenario}

In the previous sections, we have examined the complementarity and synergy offered by operating ILC even if the FCC integrated programme is the choice proposed to the CERN Council by the European Strategy. For the sake of illustration, we discuss now a complementary FCC-ee and ILC operation plan, that would improve the scientific output, in a cost-effective manner. 

As mentioned in Section~\ref{sec:accelerator}, circular colliders are the most efficient option for centre-of-mass energies below $\sim 300$\,GeV, while linear colliders progressively take over above this value. {\bf If ILC is approved by the Japanese government, \underline{and} if the assumptions leading to the claimed ILC design luminosities can be demonstrated}, it would clearly be more efficient that centre-of-mass energies around the ${\rm t \bar t}$ threshold and above (i.e., energies in excess of 340\,GeV) be covered solely by ILC, and that FCC-ee operate only at centre-of-mass energies up to the ZH cross-section maximum (i.e., energies up to 240\,GeV). To maximise the energy efficiency (quantified, e.g.,  by the number of Higgs boson per GJ), it is assumed in the following that FCC-ee operates with four interaction points, thereby multiplying the integrated luminosity per year by a factor~1.7. 

\begin{figure}[!htb]
\centering
\includegraphics[width=0.8\textwidth]{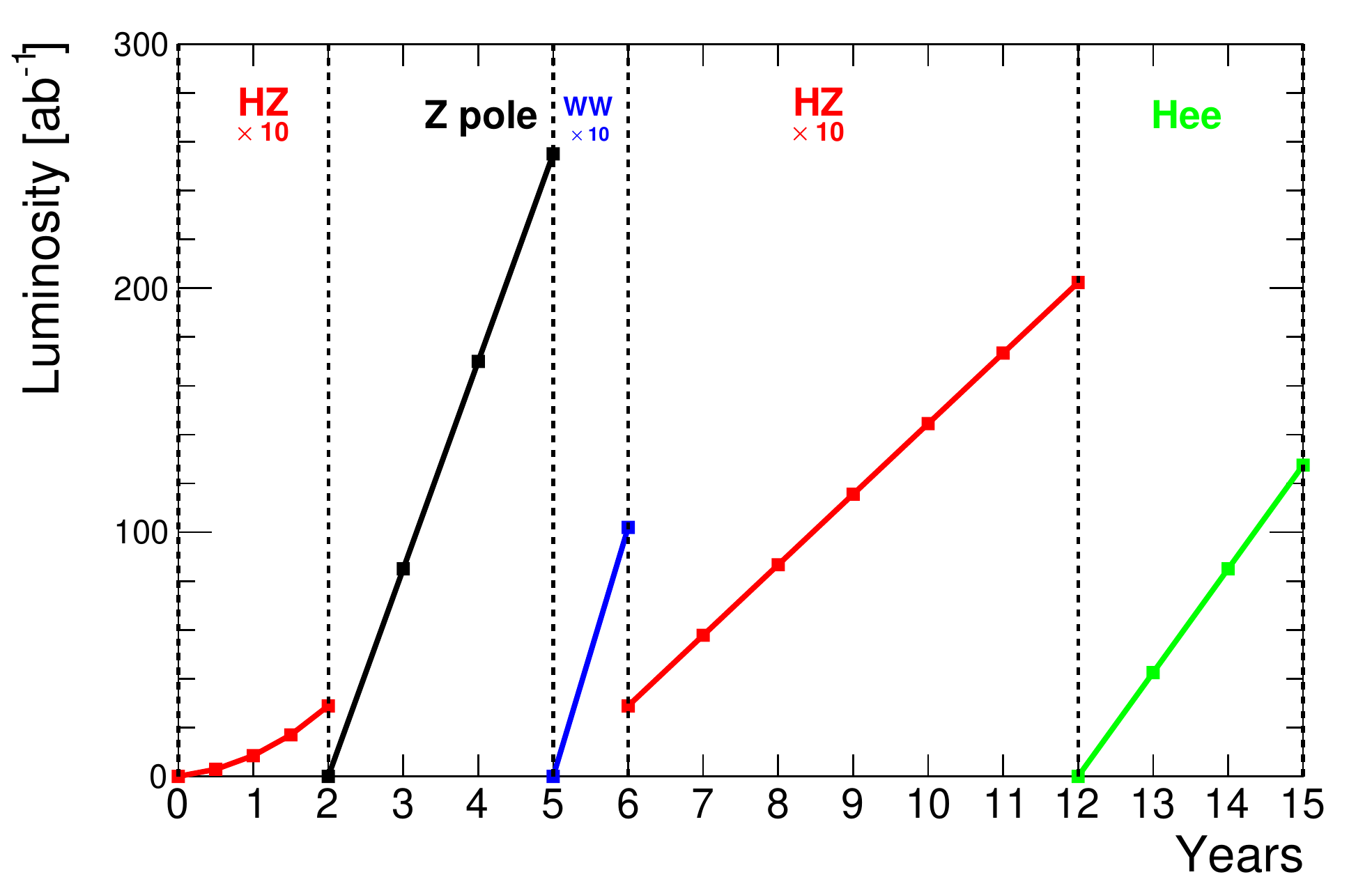}
\caption{\small 
Possible operation model of FCC-ee, complementary to ILC, without a run at the top-pair threshold and with four interaction points, over the same duration as the baseline model. In this projection, each year amounts to $1.08 \times 10^7$ seconds of operation. The order in which the energies are run can be modified, but the Hee period must be scheduled last: the scan of the Higgs resonance requires {\it (i)} the Higgs boson mass to be known to a few MeV from the 240\,GeV run; and {\it (ii)} the beam-energy calibration and stabilisation method to be fully mastered from the experience at the Z pole and the WW threshold.}
\label{fig:NewOperationModel}
\end{figure}

A timely decision from Japan would therefore be welcome in order to optimise the FCC-ee operation plan accordingly. The six years saved by not upgrading and operating FCC-ee at the ${\rm t \bar t}$ threshold and above (thus saving 1.1\,BCHF) could, for instance, be used to run three more years at 240\,GeV, and three years at $\sqrt{s} = m_{\rm H}$ with monochromatisation, so as to determine the electron Yukawa coupling. A possible FCC-ee operation plan would therefore look as shown in Fig.~\ref{fig:NewOperationModel}, starting in 2038 with two years at the ZH cross-section maximum with half the design luminosity, and accumulating $2.9\,{\rm ab}^{-1}$ (corresponding to almost 600,000 Higgs bosons) by 2040. It would then proceed with $250\,{\rm ab}^{-1}$ at and around the Z pole (almost $10^{13}$ Z produced) and $10\,{\rm ab}^{-1}$ at the WW threshold; reach $20\,{\rm ab}^{-1}$ at the ZH cross-section maximum (4 million Higgs bosons produced); and accumulate $130\,{\rm ab}^{-1}$ at the Higgs resonance. Should ILC either not be approved or deliver far short of the design performance, the top-quark precision programme at the ${\rm t \bar t}$ threshold and above must be planned with FCC-ee as suggested in Ref.~\cite{Blondel:2018aan} with typically $10\,{\rm ab}^{-1}$ at the ZH cross-section maximum and $5\,{\rm ab}^{-1}$ at 365\,GeV in view of its essential impact on the electroweak precision programme, and on the top Yukawa coupling and the Higgs WW and  self-coupling measurements.

\begin{figure}[!htb]
\centering
\includegraphics[width=0.8\textwidth]{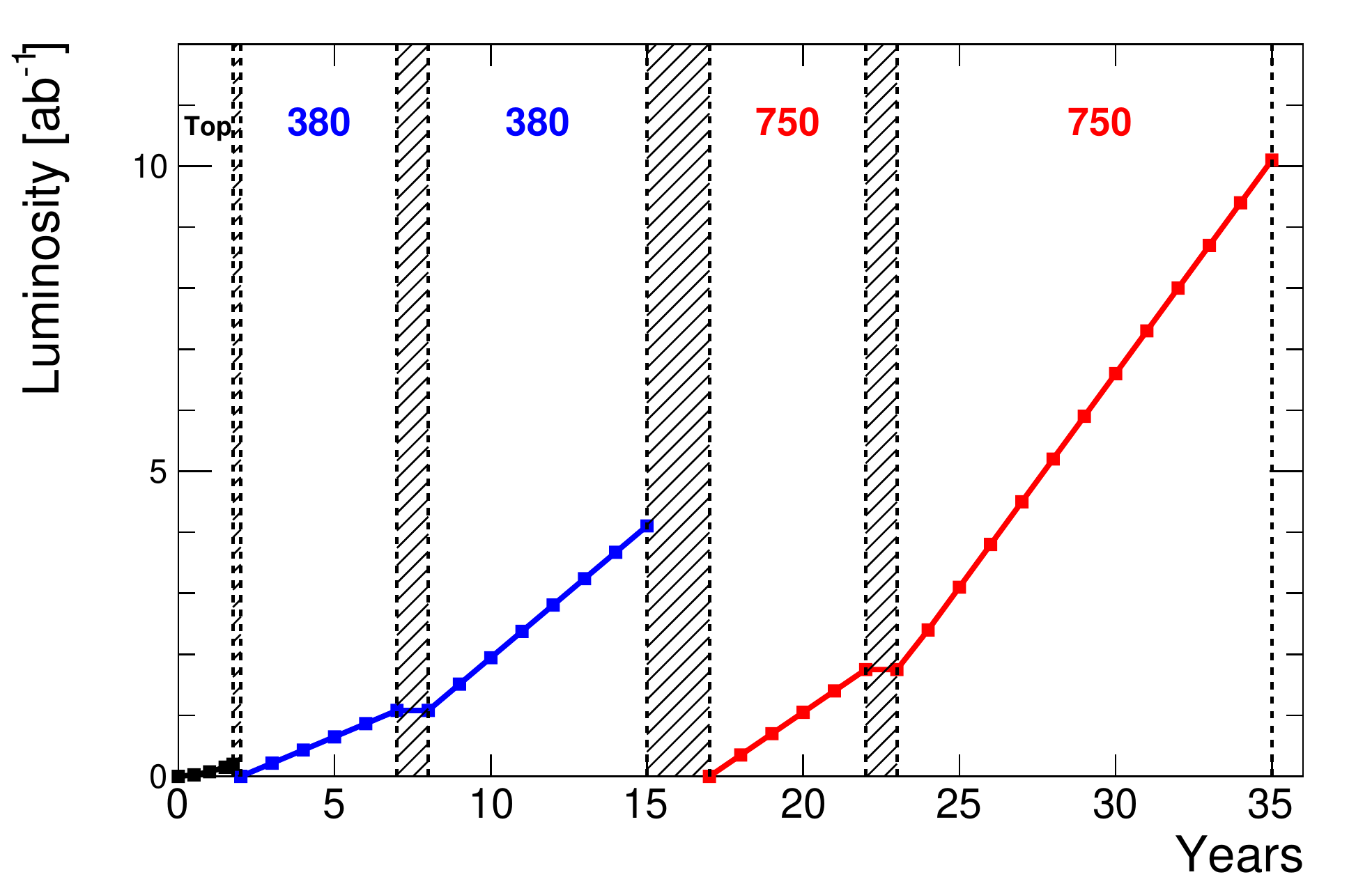}
\caption{ \small 
Possible operation model of ILC, complementary to FCC-ee, without a run at the ZH cross-section maximum, over the same duration as the baseline model. In this projection, each year amounts to $1.08 \times 10^7$ seconds of operation. The hatched areas correspond to either luminosity or energy upgrades.}
\label{fig:ILCOperationModel}
\end{figure}

If we assume -- for the sake of the exercise -- an approval of ILC in 2020, ILC operation might begin in the the early 2030's with a run at 250\,GeV to accumulate up to $2\,{\rm ab}^{-1}$ (corresponding to about 400,000 Higgs bosons) by 2045, followed by a progressive increase in energy. To maximise the scientific complementarity with the FCC-ee operation model displayed in Fig.~\ref{fig:NewOperationModel}, however, the ILC operation model should start instead with a scan of the ${\rm t \bar t}$ threshold, where 0.1 or $0.2\,{\rm ab}^{-1}$ suffice. An optimization performed by the CLIC team demonstrated that the centre-of-mass energy that offers the best complementarity for top electroweak coupling measurements and Higgs physics (also complementary to the measurements performed at the ZH cross-section maximum) is 380\,GeV. These considerations suggest that the time initially foreseen in the ILC run plan to be spent at 250\,GeV could be better invested at 380\,GeV, to accumulate the $4\,{\rm ab}^{-1}$ needed to reach the same precision on the Higgs self-coupling with single Higgs production as $5\,{\rm ab}^{-1}$ with FCC-ee at 365\,GeV. The rest of ILC programme could then be optimised for Higgs self-coupling (with double Higgs production) and top Yukawa coupling measurements. In this respect, the historical choice of 500\,GeV (which dates back to the early nineties, when neither the Higgs boson mass nor the top-quark mass were known) is not  optimal, since the cross-section maximum for ZHH production is around 600\,GeV, and the maximum for ttH production is close to 800\,GeV. In addition, both  the ILC luminosity and the ${\rm HH\nu_e\bar\nu_e}$ cross section increase with the centre-of-mass energy. All of these arguments advocate for a long run of at least $10\,{\rm ab}^{-1}$ at an  energy of $\sqrt{s} \sim 700$--$800$\,GeV. Although the 380\,GeV initial machine would be $\sim 20\%$ more costly than the 250\,GeV original proposal, the savings mentioned in Section~\ref{sec:financial} would make up for it at least partly, and the absence of a 1\,TeV upgrade would save further two-to-three billion USD with respect to the $250 \to 500 \to 1000$\,GeV baseline model. The resulting ILC operation model, complementary to that of FCC-ee in Fig.~\ref{fig:NewOperationModel}, is displayed in Fig.~\ref{fig:ILCOperationModel}, under the assumptions {\it (i)} that the run at the ${\rm t \bar t}$ operated at half the design luminosity; {\it (ii)} that all luminosity upgrades are implemented after five years of operation at each of the higher energies; and {\it (iii)} that each year amounts to $1.08 \times 10^7$ seconds.     

The scientific complementarity is illustrated in Table~\ref{tab:NewHiggsFCCILC} for the precision on the Higgs couplings obtained in the $\kappa$ framework with FCC-ee$_{240}^{\rm 4IP}$ data, ILC$_{380}$ data, and their combination. The precisions are significantly better than those obtained 
{\setlength{\tabcolsep}{6pt} 
\renewcommand{\arraystretch}{1.1} 
\begin{table}[htbp]
\centering
\caption{\small Standalone precision, in \%, on the Higgs boson couplings in the $\kappa$ framework, from a concurrent operation of  FCC-ee$_{240}$ and ILC$_{380}$, and from their combination.  For $g_{\rm HHH}$, the result of a global EFT fit is shown. Projections from HL-LHC are not included.}
\label{tab:NewHiggsFCCILC} 

\vspace{2mm}
\begin{tabular}{|l|c|c|c|}
\hline Collider & {FCC-ee$_{240}^{\rm 4IP}$} & {ILC$_{380}$} &  Combin. \\ \hline
$g_{\rm HZZ}$ (\%) & 0.11  &  0.25 & 0.11 \\ 
$g_{\rm HWW}$ (\%) & 0.65 &  0.43  & 0.31 \\ 
$g_{\rm Hbb}$ (\%) & 0.67 &  0.80  & 0.41\\ 
$g_{\rm Hcc}$ (\%) & 0.87 &  2.2  &  0.67 \\ 
$g_{\rm Hgg}$ (\%) & 0.83 &  1.2  & 0.59 \\ 
$g_{\rm H\tau\tau}$ (\%) & 0.71 &  1.3 & 0.47 \\ \hline
$g_{\rm HHH}$ (\%) & -- & --  & 24.\\ \hline
$\Gamma_{\rm H}$ (\%) & 0.9 &  1.3  & 0.8 \\ \hline
BR$_{\rm inv}$ (\%) & 0.11 &  0.33  & 0.10 \\ 
BR$_{\rm EXO}$ (\%) & 0.60 &  1.4 & 0.55 \\ \hline
\end{tabular} 
\end{table}
}
either from the FCC-ee$_{240+365}^{\rm 2IP}$ programme (Table~\ref{tab:kappaEFT}) or from the ILC$_{250+500}$ programme (Table~\ref{tab:EnergyUpgrades}) for all decays (and similar to the full FCC integrated programme for the abundant decays). The precision on the Higgs self-coupling from single Higgs production reaches 24\% from the first phase runs of the two machines, similar to what is expected from the full FCC-ee programme at 240 and 365\,GeV. The 10\% achievable from double Higgs production with ILC$_{1000}$ or CLIC$_{\rm 3\,TeV}$ can alternatively be obtained with $10\,{\rm ab}^{-1}$ at $\sqrt{s} \sim 700$--$800$\,GeV (not indicated in the Table). The run at 700--800\,GeV would also allow the top Yukawa coupling to be measured with a precision of $\sim 1.5\%$.

Proton-proton collisions at 100\,TeV with FCC-hh would complete this table with rare decays (${\rm H \to \gamma \gamma, Z\gamma, \mu^+\mu^-}$), and would improve the precision of the Higgs self-coupling and top Yukawa coupling to 5\% and better than 1\% respectively. With $10^{13}$ Z bosons produced at and around the Z pole at FCC-ee, the sensitivity of the searches for ALPs, RH neutrinos, or rare Z, b, and $\tau$ would significantly improve. With the FCC-ee run at $\sqrt{s} = 125$\,GeV, not only would the Hee coupling be measured with a precision of 15\%, but the number of light neutrino species would be determined from the ratio $\sigma({\rm e^+e^- \to \gamma \nu\bar\nu}) / \sigma({\rm e^+e^- \to \gamma \mu^+\mu^-})$ with a precision of 0.0005 or better~\cite{Abada:2019lih,Gomez-Ceballos:2013zzn}, more than one order of magnitude more accurately than the current determination~\cite{Tanabashi:2018oca,Voutsinas:2019hwu,Janot:2019oyi}. 
 
In view of the above arguments, one may be tempted to propose yet another optimization, with the circular collider programme in the LEP/LHC tunnel instead of in a new 100\,km ring. Once it is decided to build a 100 km tunnel, however, this choice is not attractive financially since much more RF acceleration is required in the small ring than in the large one. Furthermore, this choice would have paramount drawbacks. First, the luminosity is expected to be about five times smaller than FCC-ee at all energies. Second, the lack of transverse polarization above $E_{\rm beam} \sim 60$\,GeV would prevent the W mass from being measured accurately. Third, because the maximum centre-of-mass energy achievable in the LEP/LHC tunnel is not much larger than 240\,GeV, the top-quark precision programme would be  left uncovered if Japan does not, eventually, 
proceed with ILC. Therefore, the electroweak programme would be jeopardised, and both the top Yukawa coupling and Higgs self-coupling measurements of FCC-hh would be affected by large theoretical uncertainty.  Fourth, the dismantling of HL-LHC before such an ${\rm e^+e^-}$ collider can be installed would generate a long gap at CERN without collider physics. Last, but certainly not least, this ${\rm e^+e^-}$ programme in the LEP/LHC tunnel would badly lack a vibrant perspective for subsequent energy-frontier exploration in pp collisions in the same tunnel. It is therefore infinitely preferable to keep the ability to proceed with the full FCC-INT programme.

Finally, arguments similar to those presented in this note could be applied to other combinations of facilities. If we stick to the long-term goal of a long-term $\ge$ 100\,TeV pp collider at CERN (see for example Ref.~\cite{Amaldi:2019yda}), a combination of FCC-INT (at CERN) and CLIC (in Japan) would also seem very attractive.
 
 \section{Conclusions}
 \label{sec:conclusion}
 The question of complementarity between circular and linear ${\rm e^+e^-}$ colliders has been mentioned at several occasions in the past~\cite{AlainBlondel,TimBarklow,Blondel:2019yqr}. We have presented in this note a first, non-exhaustive, analysis of the complementarity of the FCC-ee and ILC physics programmes. The performance characteristics of the two machines are different. Significant domains of physics exist, in which each of these two types of machines is unique and irreplaceable. 
 \begin{itemize}
     \item For FCC-ee, the high luminosity and the exquisite energy calibration at the Z, WW, and ZH energies, and the possibility of monochromatisation at $\sqrt{s} = m_{\rm H}$, are building blocks of a unique program. The Z factory with several trillions of Z produced, offers unique opportunities for a multitude of electroweak measurements, flavour (b, c, $\tau$) physics, and searches for SM symmetry violations and feebly coupled particles. The possibility to observe the $s$-channel Higgs production leads to a unique chance of measuring the Yukawa coupling of the electron. 
     \item For ILC, the ability to explore lepton collisions above 365 GeV is unique. This is particularly interesting for the searches for new particles in the gaps possibly left by hadron colliders, and may prove essential should a new particle be discovered at the LHC in the suitable energy regime. 
 \end{itemize}
 
We have shown that this complementarity can be exploited and optimised by a suitable modification of the respective run plans, leading simultaneously to a substantial global cost saving and to an improvement of the overall physics performance, compared to the simple addition of the original run plans. The presented hypothetical scenario would obviously need, in due time, further local and global optimization by users and committees of both facilities. The practicalities and the resources needed for building both infrastructures should, of course, be studied in great detail. The following conclusions can nevertheless be drawn. 
 \begin{enumerate}
    \item Even if the European Strategy decides to support the FCC integrated programme at CERN, starting with FCC-ee towards FCC-hh as ultimate goal, the physics motivation for high-energy linear colliders remains, once their performance claims are demonstrated. Plans {and R{\&}D} for such colliders in other regions of the world should therefore be pursued.  
    \item Given the essential synergies between FCC-ee and FCC-hh, FCC-ee remains the most pragmatic, safest, and most effective way towards the 100 TeV pp exploration machine. Its physics case is extremely strong, whether or not sufficient support is found to build ILC in Japan. Altogether, it is clear that the full FCC-INT programme is a formidable tool of investigation.  
\end{enumerate}

There is no doubt that continuation of a worldwide plan including both circular and linear machines would pose considerable challenges, but it would offer the particle physics community complementary programs with a long-term vision of high-energy frontier colliders.

\bibliographystyle{jhep}
\bibliography{sample}

\end{document}